\documentclass[twocolumn]{aastex61}

\newcommand       \msun        	{$M_{\odot}$}
\newcommand       \lsun      	{$L_{\odot}$}

\newcommand	     \myr              {$M_{\odot}$~yr$^{-1}$}

\newcommand        \mic        	 {$\mu$m}

\newcommand      \spitz  {{\it Spitzer}}

\begin{document}
\title{CONSTRAINTS ON THE PROGENITOR OF SN~2010JL AND  PRE-EXISTING \\HOT DUST IN ITS SURROUNDING MEDIUM 	}
\shorttitle{Progenitor of SN 2010jl}
\author{Eli Dwek}
\affiliation{Observational Cosmology Lab, NASA Goddard Space Flight Center, Mail Code 665, Greenbelt, MD 20771, USA}
\email{eli.dwek@nasa.gov}

\author{Richard G. Arendt}
\affiliation{CRESST/UMBC/GSFC}
\affiliation{Observational Cosmology Lab, NASA Goddard Space Flight Center, Mail Code 665, Greenbelt, MD 20771, USA.}

\author{Ori D. Fox}
\affiliation{Space Telescope Science Institute, 3700 San Martin Drive, Baltimore, MD 21218, USA.}

\author{Patrick L. Kelly}
\affiliation{Department of Astronomy, University of California, Berkeley, CA 94720-3411.}

\author{Nathan Smith}
\affiliation{Steward Observatory, 933 N. Cherry Ave., Tucson, AZ 85721, USA.}

\author{Schuyler D. Van Dyk}
\affiliation{IPAC/Caltech, Mailcode 100-22, Pasadena, CA 91125, USA.}

\author{Alexei V. Filippenko}
\affiliation{Department of Astronomy, University of California, Berkeley, CA 94720-3411, USA.}
\affiliation{Senior Miller Fellow, Miller Institute for Basic Research in Science, 
University of California, Berkeley, CA 94720-3411, USA.}

\author{Jennifer Andrews}
\affiliation{Steward Observatory, 933 N. Cherry Ave., Tucson, AZ 85721, USA.}

\author{Isaac Shivvers}
\affiliation{Department of Astronomy, University of California, Berkeley, CA 94720-3411.}

\begin{abstract}
A search for the progenitor of SN~2010jl, an unusually luminous core-collapse supernova of  Type~IIn, using  pre-explosion {\it Hubble}/WFPC2 and {\it Spitzer}/IRAC images of the region, yielded  
upper limits on the UV and near-infrared (IR) fluxes from any candidate  star.  These upper limits constrain the  luminosity and effective temperature of the progenitor, the mass of any preexisting dust in its surrounding  circumstellar medium (CSM), and dust proximity to the star.  
A {\it lower} limit on the CSM dust mass is required to hide a luminous progenitor from detection by {\it Hubble}. 
{\it Upper} limits on the CSM dust mass and constraints on its proximity to the star are set by requiring that the absorbed and reradiated IR emission not exceed the IRAC upper limits. 
Using the combined extinction-IR emission constraints we present viable $M_d-R_1$ combinations, where $M_d$ and $R_1$ are the CSM dust mass and its inner radius. These depend on the CSM outer radius, dust composition and grain size, and the properties of the progenitor. The results constrain the pre-supernova evolution of the progenitor, and the nature and origin of the observed post-explosion IR emission from SN~2010jl. 
In particular, an $\eta$~Car-type progenitor will require at least 4~mag of visual extinction to avoid detection by the {\it Hubble}. This can be achieved with dust masses  $\gtrsim 10^{-3}$~\msun\ (less than the estimated 0.2-0.5~\msun\ around $\eta$~Car) which must be located at distances of $\gtrsim 10^{16}$~cm from the star to avoid detection by {\it Spitzer}.   
\end{abstract}


\section{Introduction}
Type~IIn SNe are an important  subclass of core-collapse supernovae (CCSNe) characterized by spectra that are composed of narrow, intermediate, and broad velocity line components, and luminosities that significantly exceed the power output from available radioactive elements in their ejecta \citep{schlegel90,filippenko97,smith08c}. Their spectral characteristics have been attributed to the presence of a dense ($n \gtrsim 10^8$~cm$^{-3}$) and possibly dusty circumstellar medium (CSM), presumably generated by  periods of high mass loss rates, $10^{-4}-10^{-1}$~\myr\ \citep{smith09a, fox09,fox11,moriya14}, during the pre-supernova evolution of their progenitor stars. 

The light curves of luminous Type~IIn SN are powered by the interaction of the SN  shock wave with the surrounding CSM,  and exhibit unique trends in the X-ray, ultraviolet (UV), and optical to near-infrared (NIR) continuum \citep[e.g.][and references within]{taddia13}.
Of the Type~IIn SNe, SN~2010jl stands out because it has been extensively studied at X-ray, UV, optical, and IR wavelengths \citep[][and references therein]{ofek10,smith12,maeda13,fransson14,gall14,chandra15}.   

Knowledge of the nature of the progenitor of Type~IIn SNe will provide much needed information on the stellar evolutionary process leading to such events. For SN~2010jl such knowledge can also provide important clues for the origin of its IR light curve. The progressive and wavelength-dependent blueward shift of the postshock emission lines, and the evolution of the IR light curve, have been interpreted as evidence for the rapid formation of dust in the cooling postshock region of the CSM \citep{smith12,gall14}. However, this interpretation is not unique, with IR emission from pre-existing CSM dust, heated by the SN light curve, as an alternative or additional  source of the IR emission \citep{andrews11, fransson14}. 

Early pre-explosion images of the field around SN~2010jl detected a blue source whose location is consistent with that of the supernova \citep{smith11}. More recent searches, using pre-explosion  {\it Hubble Space Telescope} ({\it HST}) Wide-Field Planetary Camera~2 (WFPC2), and {\it Spitzer} Infrared Array Camera (IRAC)  images of the region around SN~2010jl, ruled out the blue source as the progenitor, and provided upper limits on the fluxes from any such star \citep[][see Table~\ref{table:limits}]{fox17}.

\begin{deluxetable}{lcc}
\tabletypesize{\footnotesize}
\tablewidth{0pt}
\tablecaption{Upper limits on the fluxes from a SN~2010jl progenitor}
\tablehead{
\colhead{Instrument} &
\colhead{Filter (\mic)} &
\colhead{  $\nu F_{\nu}(\lambda)$\tablenotemark{1}} 
  }
 \startdata 
{\it HST} WFPC2    &  F300W 0.2892    &    0.0545     \\ 
                               &   F814W 0.8203    &   0.0563   \\
  {\it Spitzer} IRAC  &  3.6  &      0.83           \\  
                               &  4.5 &     1.67             \\ 
                               &  5.7  &     3.52             \\ 
                               &  8.0  &     6.04              \\                        
 \enddata
 \tablenotetext{1}{Fluxes in units of $10^{-17}$~W~m$^{-2}$, taken from  \cite{fox17}.}
  \label{table:limits}
\end{deluxetable}

In this paper, we use these upper limits to  constrain the  luminosity and effective temperature of the progenitor star, and the mass and inner radius of any pre-existing hot CSM dust. The UV and near-IR (UVNIR) upper limits at 0.28 and 0.82~\mic\ provide important constraints on the {\it minimum} CSM dust mass and the proximity of the CSM to the progenitor star in order for its emission not to exceed the observed upper limits. The IRAC 3.6--8.0~\mic\ upper limits provide useful {\it upper} limits on the dust mass and proximity of the CSM, since too much dust located too close to the progenitor star will give rise to an IR flux that will exceed the observed upper limits. 

We first describe the general  assumptions and parameters of our model (Section 2). In Section~3 we derive upper limits on the line-of-sight (LOS) extinction due to Galactic or host-galaxy dust. The minimum mass and maximum CSM dust masses required to hide the progenitor star without exceeding the IRAC upper limits are derived in Section~4. Limits on the dust mass, composition, and grain sizes as a function of progenitor and CSM properties are presented in Section~5. The results of the paper are presented in Section~6. 

\section{Model assumptions and parameters}
Possible SNIIn progenitors are massive luminous stars with large mass-loss rates, including  hot (20,000~K) and cool (10,000~K) luminous blue variables (LBVs), and yellow and red supergiant stars (referred to as YSG and RSG stars, respectively). 
In our model we explore the stellar luminosities and temperatures from $10^4$ to $10^7$~\lsun, and $10^3$ to $10^5$~K, respectively, and present detailed results for progenitors with luminosities and temperatures listed in Table~\ref{table:prog}.

\begin{deluxetable}{lcl}
\tabletypesize{\footnotesize}
\tablewidth{0pt}
\tablecaption{Physical parameters of progenitor types\tablenotemark{1}}
\tablehead{
\colhead{Progenitor} &
\colhead{Temperature (K)} &
 \colhead{Luminosities (\lsun)} 
  }
 \startdata 
 Hot LBV  & 22,000   &      $1\times 10^5,~5\times 10^5, ~1\times 10^6, ~5\times 10^6  $ \\ 
 Cool LBV & 10,000  &      $1\times 10^5, ~5\times 10^5, ~1\times 10^6, ~5\times 10^6  $ \\ 
 YSG        & 5,600    &      $1\times 10^5, ~2.5\times 10^5, ~6.3\times 10^5                   $ \\
 RSG        & 3,500    &      $1\times 10^5, ~2.5\times 10^5, ~6.3\times 10^5                   $\\
 \enddata
  \tablenotetext{1}{LBV = Luminous blue variable, YSG = Yellow supergiant, \\ RSG = red supergiant}
  \label{table:prog}
\end{deluxetable}

We assumed that the CSM formed during a period of steady mass loss from the progenitor, and that consequently the gas has an $r^{-2}$ density profile. We further assumed a constant dust-to-gas mass ratio in the outflow, so that the dust follows the same density profile as the gas.  
We studied the extinction and emission constraints as a function of CSM geometry, the inner, $R_1$, and outer, $R_2$, radius of the CSM shell, and its total dust mass, $M_d$. In spite of evidence that the CSM shell may be narrow, and may extend only to radii $\lesssim 10^{17}$~cm \citep{ofek14,chandra15}, we explored a wide range of $R_1$ values, from $10^{15}$~cm, approximately the value of the breakout radius \citep{ofek14}, to a value of $10^{18}$~cm. $R_2/R_1$ ratios, expressed in terms of $R_2/R_1=\xi^{-1}$ throughout the paper, were taken to be 1.2 (thin shell), 3, and 10 (extended shell). We took CSM dust masses ranging from $10^{-6}$ to 0.1~\msun, the largest mass expected from a 10~\msun\ shell with about a solar abundance of metals.  

For such density profile, most of the emission that is constrained by the IRAC fluxes emanates from hot dust located near $R_1$. To constrain the mass of cold dust in the CSM, we constructed a simple radiative transfer model (Sec. 4.2), in which we calculated the dust temperature as a function of distance from the progenitor star. Consequently, the mass of cold dust is related to that of the hot dust at radius  at $R_1$. All reported upper limits on the dust mass imposed by the IRAC observations therefore include all the dust mass contained in the CSM shell, excluding any possible mass that deviates from the assumed $r^{-2}$ density profile.  

Constraints on the nature of the progenitor star and the CSM dust depend on the dust extinction and emission efficiencies, which are determined by the dust composition and size distribution. \cite{williams15} attempted to characterize the composition of dust in the CSM using post-explosion observations obtained by archival \spitz\ 3.6 and 4.5~\mic\ SN survey data, and  by 11.1~\mic\ observations obtained by SOFIA ({\it Stratospheric Observatory for Infrared Astronomy}). Any hot silicate dust giving rise to the observed 3.6 and 4.5~\mic\ fluxes would produce a large 11.1~\mic\ flux stemming from the 9.7~\mic\ silicate feature.  The SOFIA observations yielded only an upper limit, from which \cite{williams15}  ruled out silicates as a viable dust component in an optically thin (at mid-IR wavelengths) CSM. However, they conceded that optical depth effects may conceal the true nature of the radiating dust, a suggestion that we explore in more detail below. 

\begin{figure*}[t]
\begin{center}
\includegraphics[width=3.2in]{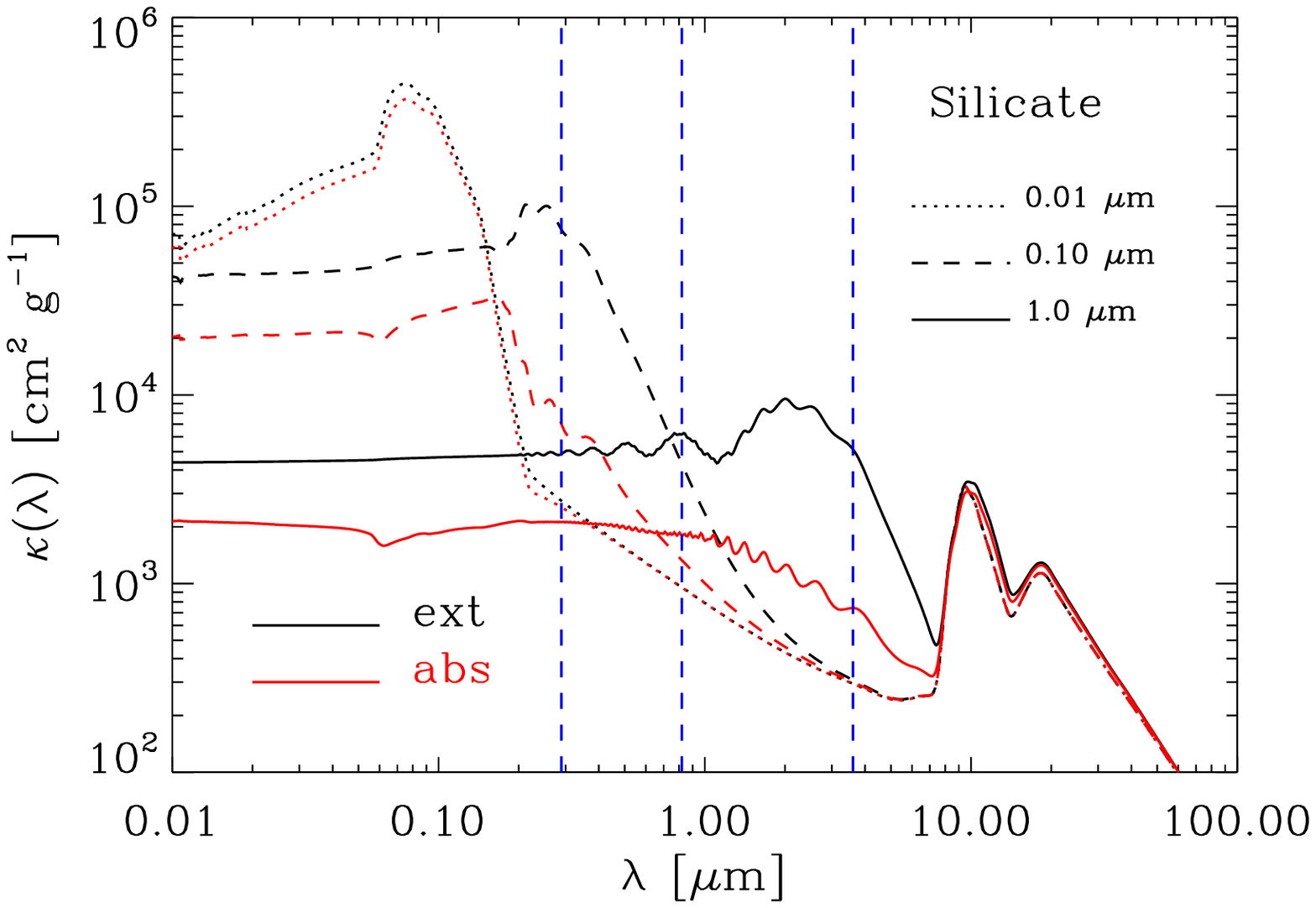}
\includegraphics[width=3.2in]{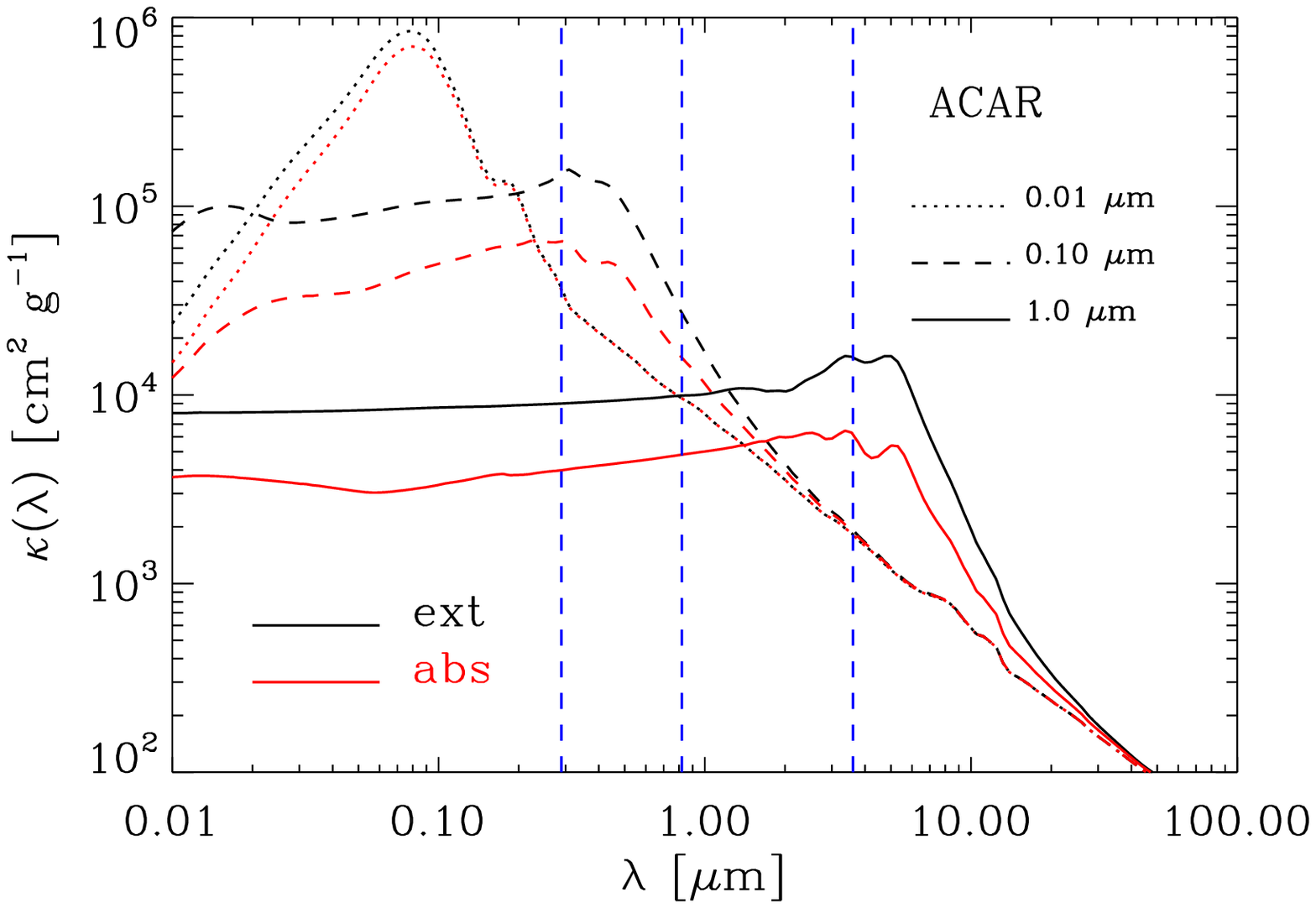}
\caption{\label{kappa} The mass absorption and extinction coefficients of the silicate (left panel) and amorphous carbon (right panel) dust grains used in this paper.  Dotted, dashed, and solid lines correspond to grain sizes of  0.01, 0.1, and 1.0~\mic\  radii grains, respectively. Black and red curves represent the mass extinction and absorption coefficients. The vertical dashed lines indicate the central wavelengths of the WFPC2 F300W and F814W bands and the IRAC 3.6 \mic\ band. 
}
\end{center}
\end{figure*}

We will therefore present our results for the two most common dust compositions: astronomical silicates and amorphous carbon (ACAR). We considered three distinct grain radii, $a$, of 0.01, 0.1, and 1.0~\mic. Figure~\ref{kappa} plots the value of the dust mass extinction and absorption coefficient, $\kappa_{ext}(a,\lambda)$ and $\kappa_{abs}(a,\lambda)$, as a function of wavelength for different grain sizes, where
\begin{equation}
\label{eq_kappa}
\kappa_{ext}(a,\lambda) = {\pi a^2\, Q_{ext}(a,\lambda) \over m_d(a)} \qquad ;
\end{equation}
 $m_d$ is the grain's mass, and $Q_{ext}$ its  extinction  coefficient. An identical equation defines $\kappa_{abs}$. Optical constants for the silicate grains were taken from \cite{weingartner01}, and for the ACAR grains, of AC1 form, from \cite{rouleau91}.   

Finally, we adopted a distance of 50~Mpc to UGC~5189A, the host galaxy of SN~2010jl \citep{smith11}.

\section{Line-of-Sight Extinction to the Progenitor star}
Some of the extinction to the progenitor could be due to dust residing in the interstellar medium of the Milky Way (MW) and the SN host galaxy. 
From the extinction maps of \cite{schlafly11}  we estimate a value of $A_{\rm V} = 0.075$ mag through the MW. 

Extinction estimates can also be obtained from measurements of the equivalent width (EW)  of the diffuse interstellar band (DIB) feature at 5780 \AA, and  the  EW of the Na\,{\small I}\, D doublet 5890, 5896 \AA\ lines. The carriers of the DIBs are still unknown; however, their EW is linearly correlated with the amount of visual extinction \citep{phillips13}. Correlation plots between the EW of the Na\, D bands and $E$(B-V)  were presented by \cite{poznanski12} and \cite{phillips13}.

We analyzed a moderate-resolution spectrum of SN~2010jl taken with the DEIMOS spectrograph \citep{faber03} mounted on the Keck-II 10\,m telescope with the 1200~line~mm$^{-1}$ grating on November 7, 2010 UT, approximately four days after discovery \citep{smith12}. The lack of any absorption in the DIB band gives a 3$\sigma$ upper limit on the EW of 0.03 \AA\ which, using Eq.~6 of \cite{phillips13}, gives a 3$\sigma$ upper limit of $A_{\rm V} \lesssim 0.7$\,mag. 

From the DEIMOS spectrum we also measured a Na\,D EW  of 0.107~\AA\ in the MW, and an EW of 0.18 \AA\ (D$_1$ + D$_2$) at the  host-galaxy redshift. Assuming a value of $R_{\rm V} = 3.1$ we get  a value of $A_{\rm V}=0.1\pm0.04$~mag for the MW, consistent with that derived from the \cite{schlafly11} map, and a value of $A_{\rm V}=0.15\pm0.07$~mag for the host-galaxy, giving a total value of  $A_{\rm V} = 0.25\pm0.08$~mag.

Part of the Na\,D absorption could have occurred in the CSM of the progenitor of SN 2010jl. The CSM environment may be warmer than that of the general ISM of the MW or the host galaxy, exciting the Na and lowering the EW of the doublet line, thereby invalidating the MW-calibrated relations of \cite{phillips13}.  In the extreme case, Na\,D absorption could take place in a dust-free CSM if the progenitor did not form any dust in the outflow.

We also analyzed an SDSS spectrum 2.8$''$ from the position SN 2010jl, and measured a value of  $A_V = 0.62 \pm 0.024$~mag from the Balmer decrement along the LOS. The relevance of this value to the LOS extinction to the SN is unclear, since the extinction could vary significantly along these different sight lines. 

Considering the uncertainties in the CSM contribution to the Na\,D absorption, and the fact that the Balmer decrement measurements were offset from the LOS to the SN, we  will simply use the lack of detection of any DIB absorption to set a strict 3$\sigma$ upper limit of 0.7~mag, or $\tau_{\small LOS}(V) \lesssim 0.64$, on the visual LOS extinction through the ISM of the MW and the host galaxy. 
 
\section{Extinction  and Emission Constraints from UVNIR Limits}
\subsection{Minimum Extinguishing Dust Mass Derived From UVNIR Upper Limits}

\begin{figure*}[t]
\begin{center}
\includegraphics[width=2.3in]{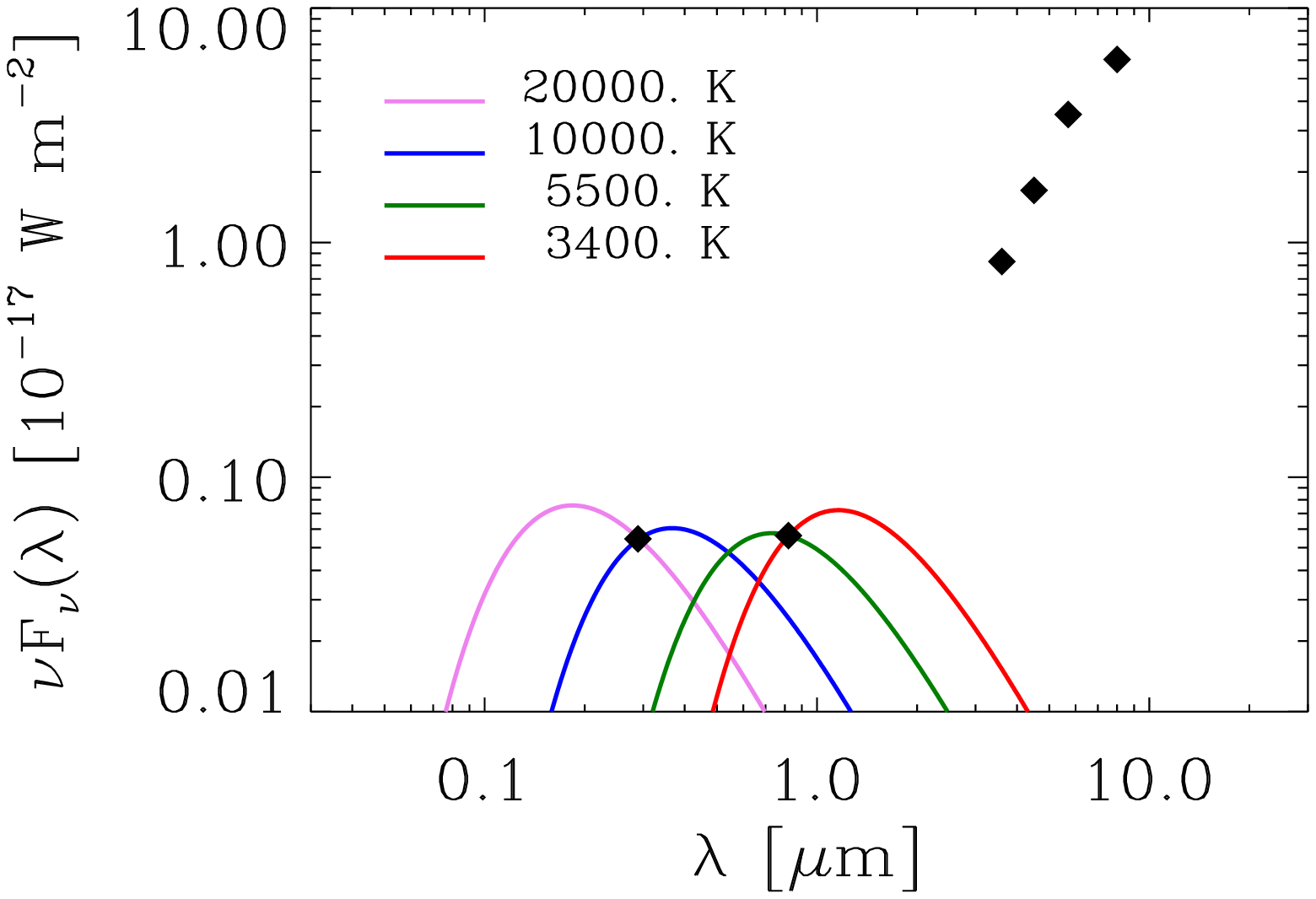}
\includegraphics[width=2.3in]{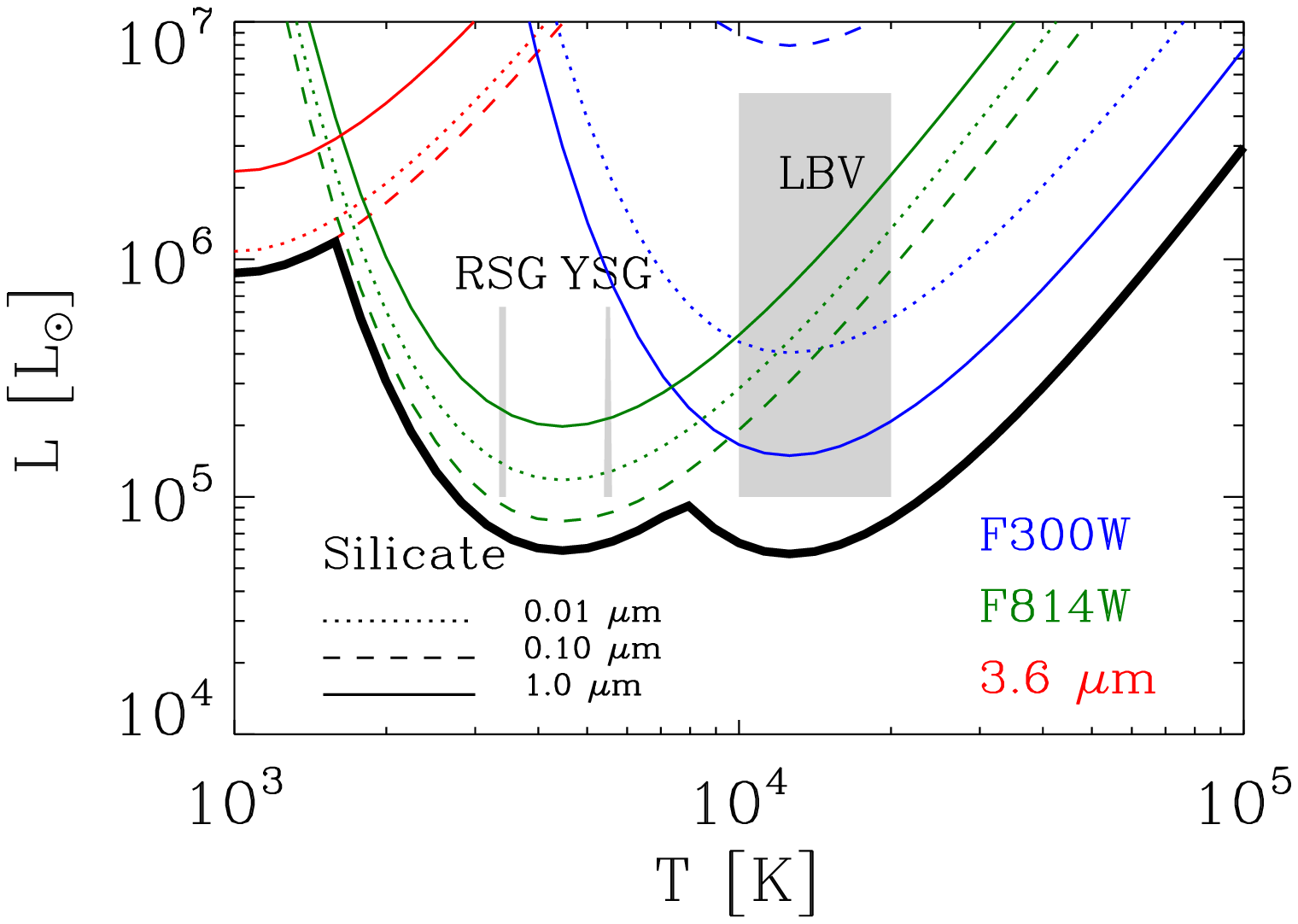}
\includegraphics[width=2.3in]{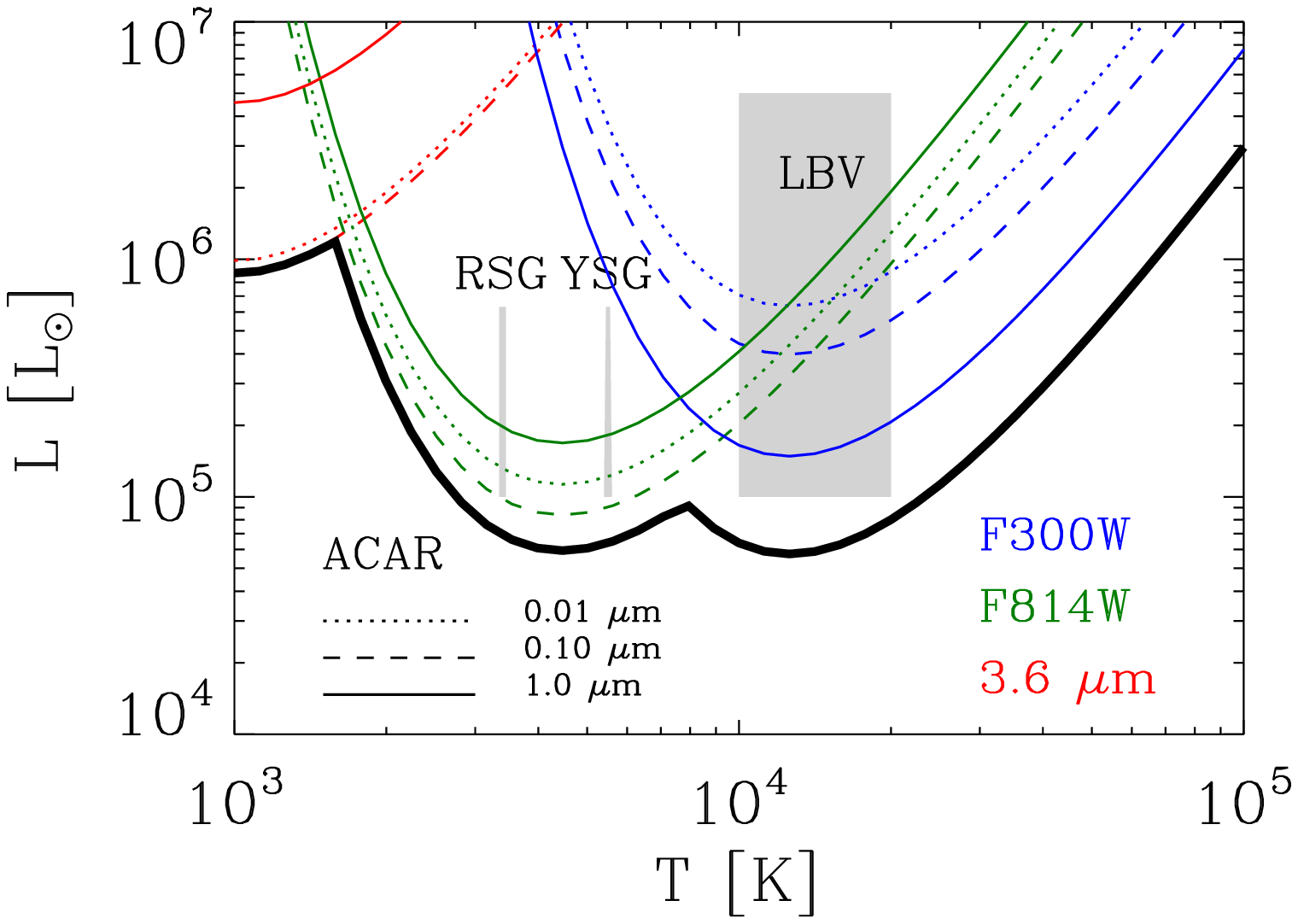}\
\caption{\label{upplim1} {\it Left panel}: Blackbody spectra characterized by temperatures of $2\times 10^4$~K (pink), $1\times 10^4$~K (blue), $5.5\times 10^3$~K (green), and $3.4\times 10^3$~K (red) are normalized to the UVNIR upper limits from the progenitor star. In the absence of extinction, the luminosity of hot progenitor stars (T $\gtrsim 8,000$~K) is constrained by the UV upper limit, whereas that of colder stars is constrained by the NIR upper limit. For the 50~Mpc distance to SN~2010jl, $1\times10^{-17}$~W~m$^{-2}$ corresponds to a luminosity of $7.8\times10^5$~\lsun.  {\it Middle panel}:  
The black line is the upper limit, in the absence of extinction, on the progenitor luminosity as a function of stellar temperature, as imposed by the observational limits in the left panel. If extinction corresponding to $\tau_V = 1$ is present, the upper limit on the luminosity rises to the curves indicated by the solid, dashed, or dotted lines for grain sizes of 1, 0.1, or 0.01 \mic, respectively. The red, green, and blue colors indicate the luminosity constraint for each band separately. Gray regions mark the locations of RSGs, YSGs, and LBVs.
}
\end{center}
\end{figure*}

The upper limits on the UVNIR emission from the position of the progenitor provide important constraints on the minimum amount of extinction required to hide a massive star progenitor.  The specific flux, $F_{\nu}(\lambda)$, at wavelength $\lambda$  from a star located at distance $D$, and characterized by a given luminosity $L_s$ and a blackbody spectrum of temperature $T_s$ is given by
\begin{equation}
\label{eq_fnu}
F_{\nu}(\lambda)=\left({L_s\over 4\pi D^2}\right)\, \left({\pi B_{\nu}(\lambda,T_s)\over \sigma T_s^4}\right) \ .
\end{equation}
Figure~\ref{upplim1} (left panel) shows the flux, $\nu\, F_{\nu}(\lambda)$, of several  stars, with blackbody temperatures of  20,000, 10,000, 5,500, and 3,400~K, whose luminosities were normalized to match the observed UVNIR upper limits. 

Stars with  higher luminosities require a minimum amount of  extinction, given by
\begin{equation}
\label{eq_tau}
\tau_{ext}^{\rm min}(\lambda) =  \ln \left[ {F_{\nu}(\lambda) \over F_{\nu}^{{\rm up}}(\lambda)}\right] \ , 
\end{equation}
to drop their fluxes below the observed upper limits, where $F_{\nu}^{{\rm up}}(\lambda)$ is the observed upper limit  on the flux from the progenitor star.

The middle and right panels in Figure \ref{upplim1}  show the constraints on the luminosities and blackbody temperatures of candidate progenitors of SN~2010jl. The $L_s-T_s$ parameter space encompasses that spanned by RSGs, YSGs and LBVs. Their adopted luminosities and effective blackbody temperatures are listed in Table~2. The grey areas in the figure mark their locations in the $L_s-T_s$ diagram. 

The thick black curve represents the upper limit on the stellar luminosity as a function of stellar temperature in the absence of any extinction.  It clearly shows the temperature ranges in which the various observed upper limits dominate the constraints: the IRAC 3.6~\mic\ upper limit for stars with $T \lesssim 1,800$~K, the F814W upper limit for stars with $1,800 \lesssim T(K) \lesssim 8,000$, and the F300W upper limit for hotter stars. 
These temperature ranges  change in the presence of extinction.
The blue, green, and red curves show the effect of a $\tau(V)=1$ extinction on the upper limits on the progenitors' luminosity as a function of temperature. The colors correspond to the band providing the constraint. The magnitude of the increase in the upper limit on the stellar luminosities depends on the grain size, since for a fixed extinction of unity at $V$, the extinction in the other bands is set by the wavelength dependency of the extinction curve (Figure~\ref{kappa}). 

Not all progenitors require a dusty CSM to be hidden at UVNIR wavelengths. 
Since the observations consist of several wavelengths, the minimum  extinction required to hide a progenitor is determined by the wavelength for which $\tau_{ext}^{min}(\lambda)$ has the greatest value. For each dust composition and grain size, the extinction at that wavelength, $\lambda_m$, can be converted to a visual extinction:
\begin{equation}
\label{tauV_ext}
\tau_{ext}^{\rm min}(V)=\tau_{ext}^{\rm min}(\lambda_m)\, {\kappa_{ext}(a,V)\over\kappa_{ext}(a,\lambda_m)} \ . 
\end{equation}
The ratio $\kappa_{ext}(a, \lambda_m)/\kappa_{ext}(a,V)$ is a normalized dust extinction law for a given grain radius.  

\begin{figure}[t]
\begin{center} 
\includegraphics[width=3.0in]{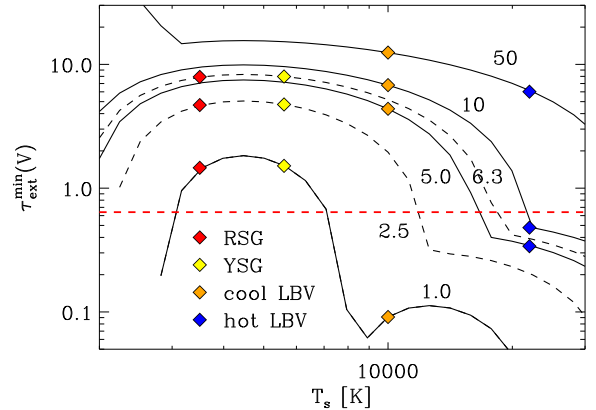}   
\caption{\label{tauv} The visual optical depth, produced by 0.1~\mic\ silicate grains,  required to hide a star [see Eq. (\ref{eq_tau})] as a function of stellar temperature for different luminosities of (from bottom to top) $1.0\times10^5$, $2.5\times10^5$, $5.0\times10^5$, $6.3\times10^5$, $1\times10^6$, and $5\times10^6$~\lsun. Diamonds indicate the location of the hot LBV (blue), cool LBV (orange), YSG (yellow), and RGB (red) in the $L_s-T_s$ diagram. The horizontal red dashed line marks the 3$\sigma$ upper limit on the LOS optical depth (MW+host galaxy) of $\tau(V)=0.64$ to the progenitor star. 
}
\end{center}
\end{figure}

Figure~\ref{tauv} depicts the visual optical depth, given by Equation~(\ref{tauV_ext}), required to hide a progenitor star as a function of stellar temperature for different luminosities. The diamonds mark the locations of the various progenitors (see Table~2) in the figure. The horizontal red dashed line represents the 3$\sigma$ upper limit on the LOS extinction provided by the Milky Way and the host galaxy. Progenitors that have an extinction below this value may not require a dusty CSM. 

The minimum amount of visual extinction required to obscure a given progenitor star  can be translated into a minimum dust mass. 
We will assume that the CSM was created in a steady spherically symmetric outflow, so that the dust density, $\rho_d(r)$, at a given radius $r$  is given by
\begin{eqnarray}
\label{rhod}
\rho_d(r) & = & \int_{a_1}^{a_2}\, \tilde{\rho}_d(a,r)\, {\rm d}a \\ \nonumber
    & = & \rho_0\, \left({r\over R_1}\right)^{-2}  \qquad {\rm for} \ \ R_1 \leqslant r \leqslant R_2\\ \nonumber
          & = & 0  \qquad \qquad  \qquad \ \  {\rm for} \ \  r < R_1, \ r>R_2 \ ,
\end{eqnarray}
where $\tilde{\rho}_d(a,r)\, $d$a$ is the density of dust grains in the $a-a+$d$a$ radius interval, given by
\begin{equation}
\label{ }
\tilde{\rho}_d(a,r) = \rho_0\, {m_d(a)\over \left<m_d\right>}\, f(a) \ , 
\end{equation}
where $f(a)$ is the grain size distribution  normalized to unity in the \{$a_1, a_2$\} radius interval, and  $\left<m_d\right>$ is the average grain mass.

The total mass of dust, $M_d$, in the CSM is given by
\begin{equation}
\label{mdtot}
M_d = \int_{R_1}^{R_2} \ 4\, \pi \, r^2 \rho_d(r) {\rm d}r = {4 \pi\over \xi}\, \rho_0\, R_1^3\, (1-\xi)\ ,
\end{equation}
where $  \xi = R_1/ R_2$.

The extinction opacity out to a radius $r$, $\tau(\lambda,r)$, is
\begin{eqnarray}
\label{taudr}
\tau_{ext}(\lambda,R_1,r) & = & \int_{a_1}^{a_2}\, {\rm d}a\, \kappa_{ext}(a,\lambda)\, \ \int_{R_1}^r\ \tilde{\rho}_d(a,r)\, \, {\rm d}r   \\ \nonumber
   & = & \tau_0(\lambda,R_1)\, \left({r-R_1\over r}\right) \ ,
\end{eqnarray}
where 
\begin{equation}
\label{ }
\tau_0(\lambda,R_1) = \rho_0\, \left<\kappa_{ext}(\lambda)\right> \, R_1 \ ,
\end{equation}
 and $\left<\kappa(\lambda)\right>$ is the size-averaged dust mass extinction coefficient given by
\begin{equation}
\label{kappa_av}
\left<\kappa_{ext}(\lambda)\right> \equiv \int_{a_1}^{a_2}\ {m_d(a)\over \left<m_d\right>}\, \kappa_{ext}(a,\lambda)\, f(a)\, {\rm d}a  \ .
\end{equation}
 The total optical depth through the CSM is given by
\begin{equation}
\label{tauR}
 \tau_{ext}(\lambda,R_1,R_2) = \tau_0(\lambda,R_1)\, (1-\xi) \ .
\end{equation} 

The extinction optical depth out to radius $r$ can  be expressed in terms of the total dust mass
\begin{equation}
\label{ }
\tau_{ext}(\lambda, R_1,r) = {\xi \over 4(1-\xi)}\ \left({r-R_1\over r}\right)\, \left[{M_d\over \pi R_1^2}\right]\, \left<\kappa_{ext}(\lambda)\right> \qquad, 
\end{equation}
and its total value is
\begin{equation}
\label{tauext}
\tau_{ext}(\lambda,R_1,R_2)= {\xi\over 4}\, \left({M_d\over \pi R_1^2}\right)\, \left<\kappa_{ext}(\lambda)\right> \ .
\end{equation}
 
In all calculations we will assume that $f(a)$ is given by a delta~function at grain radii of 0.01, 0.1, and 1.0~\mic, so that $\left<\kappa_{ext}(a,\lambda)\right>$ is simply the mass extinction coefficient of a grain with radius $a$. 
Using Eqs.~(\ref{tauext}) and (\ref{tauV_ext}), the minimum dust mass in the shell can be written as
\begin{eqnarray}
\label{mdmin}
M_d^{\rm min}( R_1,R_2)  & = & {4 \pi R_1^2\over \xi}\ \left[{\tau_{ext}^{\rm min}(V) \over \kappa_{ext}(a,V)}\right] \\ \nonumber
&=  &{6.3\times 10^{-5} R_{16}^2\over \xi}\, \left[{\tau_{ext}^{\rm min}(V) \over \kappa_{ext,4}(a,V)}\right] \ \  M_{\odot} \ ,
\end{eqnarray}  
where $\kappa_{ext,4}(V)$ is the dust mass absorption coefficient in the $V$ band normalized to a value of $10^4$~cm$^2$~g$^{-1}$, and $R_{16}$ is equal to $R_1$ in units of $10^{16}$~cm.
 
The minimum mass of CSM dust required to hide the progenitor star is reduced if some of the extinction  is along the LOS to the SN. The minimum dust mass is then given by
\begin{equation}
\label{mdminLOS}
M_d^{\rm min}( R_1,R_2)   =  {4 \pi R_1^2\over \xi}\ \left[{\tau_{ext}^{\rm min}(V) -\tau_{\small LOS}(V)\over \kappa_{ext}(a,V)}\right]  \ ,
\end{equation}  
 where $\tau_{\small{LOS}}(V)$ is the LOS extinction.
\begin{figure*}[t]
\begin{center}
\includegraphics[width=3.2in]{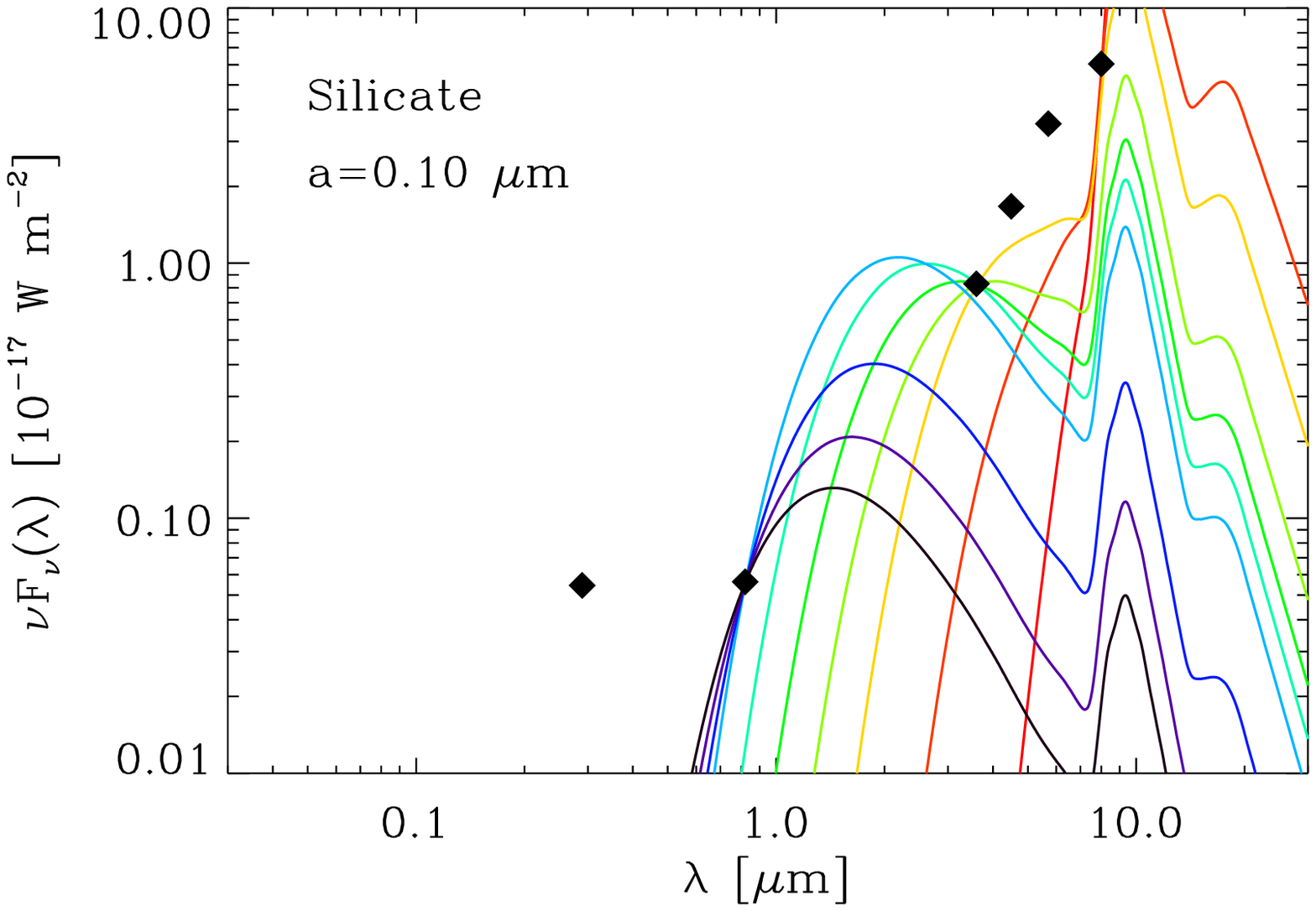}
\includegraphics[width=3.2in]{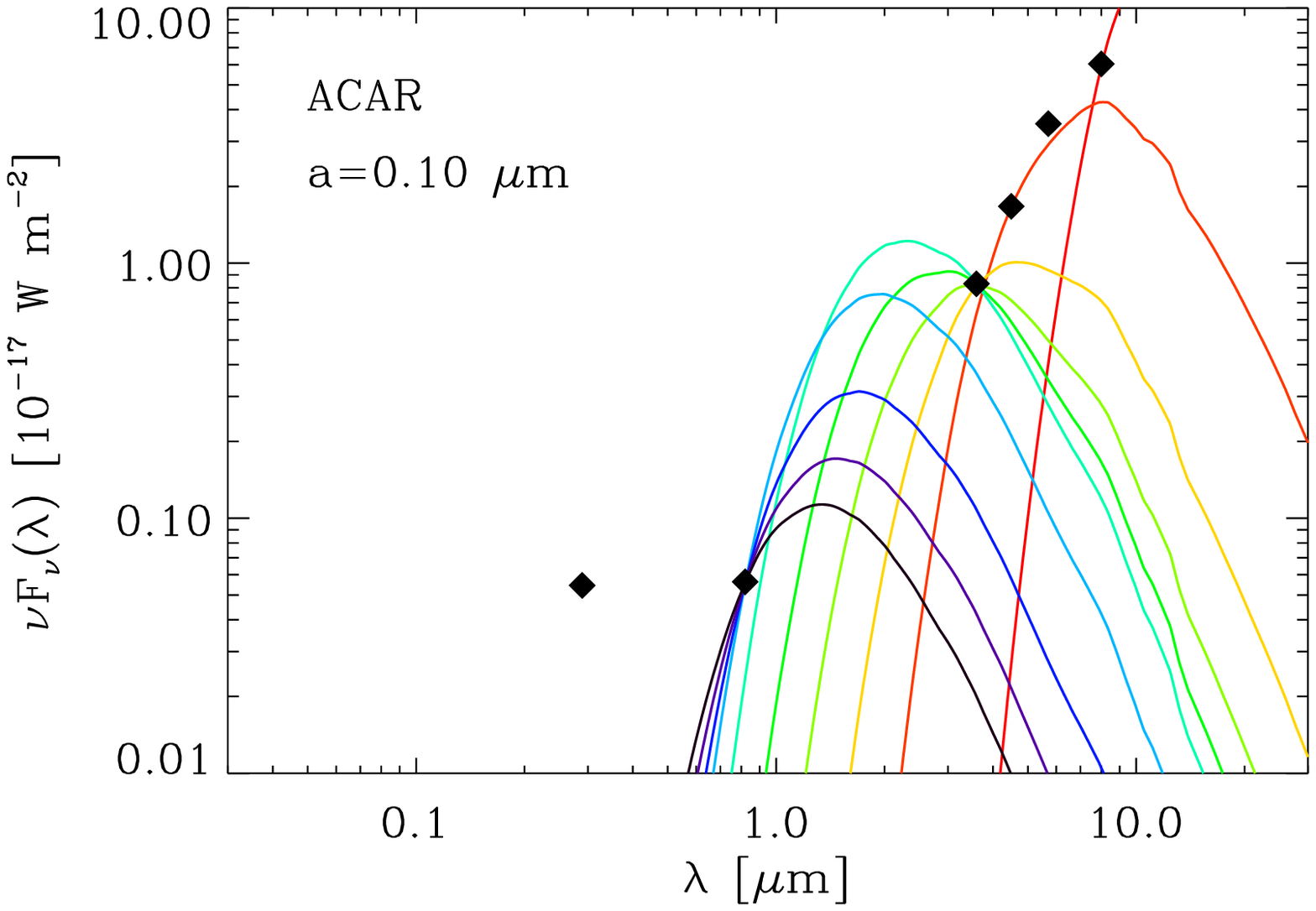}
\caption{\label{upplim2} {\it Left panel}: Possible 0.10~\mic\ radii silicate dust spectra constrained by the {\it HST} and {\it Spitzer} upper limits. {\it Right panel}: The same as the left panel for ACAR grains.  
}
\end{center}
\end{figure*}

\subsection{Maximum Emitting Dust Mass from IR Upper Limits}
 
Figure~\ref{upplim2}  presents the upper limits on the 3.6 to 8.0~\mic\ emission from any preexisting dust in the CSM.
The curves in the figure correspond to select silicate and ACAR dust spectra characterized by a range of dust temperature-mass combinations that comply with the observed IRAC upper limits.
These upper limits constrain  the dust temperature and IR emission from the CSM. Given a flux and dust temperature one can readily calculate an upper limit on the mass of the dust that is consistent with the observed upper limit.  

\begin{figure*}[t]
\begin{center}
\includegraphics[width=3.2in]{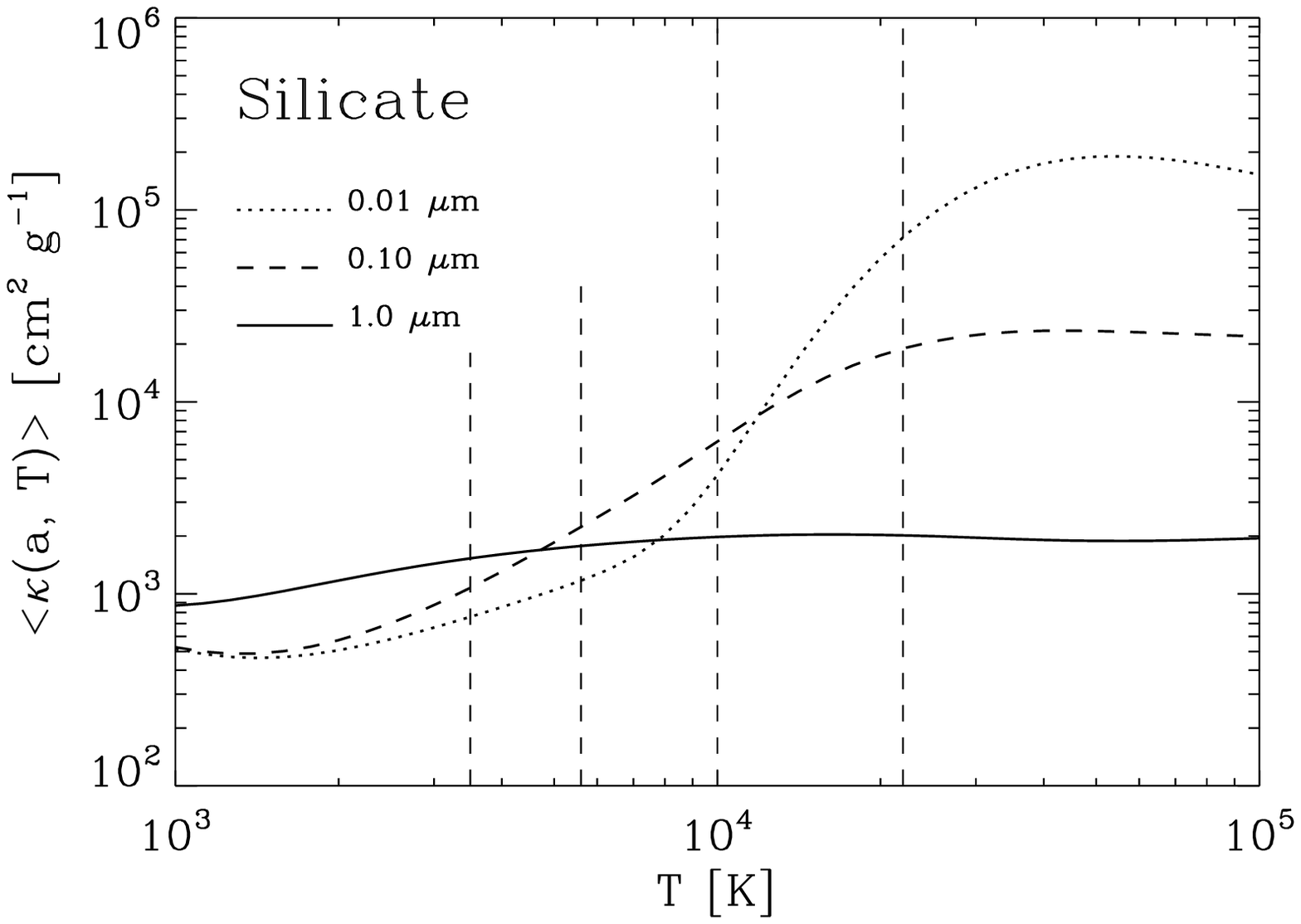}
\includegraphics[width=3.2in]{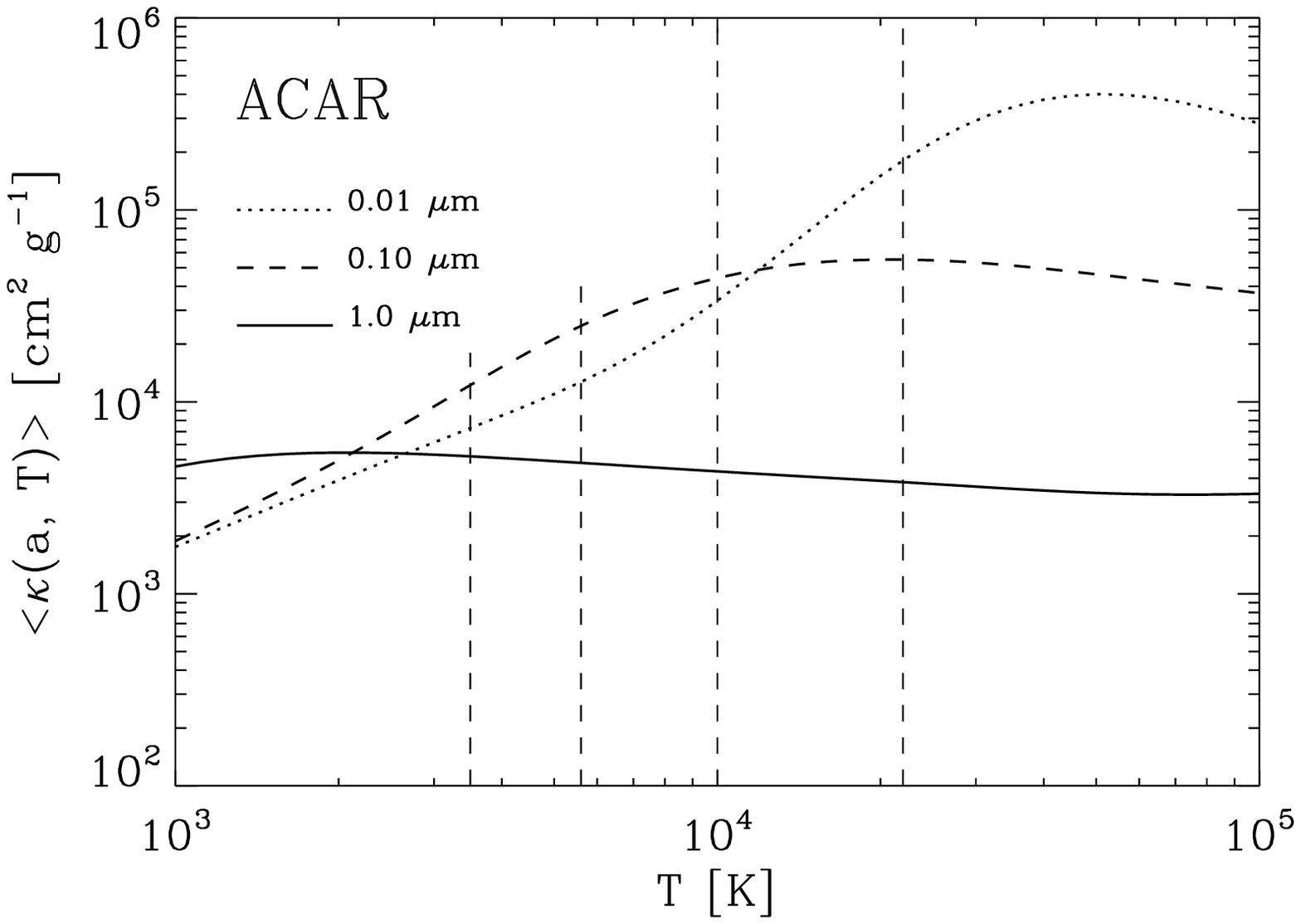}
\caption{\label{kappaTemp} The Planck-averaged value of the dust mass absorption coefficient given by Eq.~(\ref {kappaT}) is plotted agains stellar temperature for silicate (left panel) and ACAR grains (right panel) of different radii. The vertical dashed lines indicate the  temperatures of the progenitor stars listed in Table~\ref{table:prog}.}
\end{center}
\end{figure*}

The specific IR luminosity emitted by the CSM dust is given by
\begin{eqnarray}
\label{ }
L^d_{\nu}(\lambda) & = & 4\, \int \, {\rm d}m_d(r)\, \kappa_{abs}(a,\lambda)\, \pi B_{\nu}[\lambda, T_d(r)] \\ \nonumber
 & = & 4\, \int_{R_1}^{R_2}\ 4\pi\, r^2\, \rho_d(r)\, \kappa_{abs}(a,\lambda)\, \pi B_{\nu}[\lambda, T_d(r)] \, {\rm d}r \\
 & = & {4 \over 1-\xi}\, M_d\ \int_{\xi}^1\ \kappa_{abs}(a,\lambda)\, \pi B_{\nu}[\lambda, T_d(x)]\, {\rm d}x \nonumber \ ,
\end{eqnarray}
where $x\equiv r/R_2$,  and $T_d(r)$ is the dust temperature at distance $r$, derived by equating the dust heating rate by the stellar radiation to its IR cooling rate
\begin{eqnarray}
\label{Tdr}
{L_s \over 4\, \pi\, r^2}\, \int_0^{\infty} {\pi B_{\nu}(\lambda,T_s)\, \exp[-\tau(\lambda,r)] \over \sigma T_s^4}\, \kappa_{abs}(a,\lambda) {\rm d}\nu &  & \\ \nonumber
 = 4\, \sigma\, T_d(r)^4\, \left<\kappa_{abs}(a,T_d(r))\right>  & & .
\end{eqnarray}
For any temperature $T$, $\left<\kappa_{abs}(a,T)\right>$ is defined as
\begin{equation}
\label{kappaT}
\left< \kappa_{abs}(a,T)\right> \equiv  { \int_0^{\infty}\, \pi B_{\nu}(\lambda,T)\, \kappa_{abs}(a,\lambda) \, d\nu  \over  \sigma T^4} \ .
\end{equation} 
Figure~\ref{kappaTemp} shows the behavior of $\left<\kappa_{abs}(a,T)\right>$ as a function of temperature.  

In particular, the dust temperature, $T_d(R_1)$, at the inner boundary of the CSM shell is given by 
\begin{equation}
\label{Td1}
 \left({L \over 16\, \pi\, R_1^2}\right)\,  \left<\kappa_{abs}(a,T_s)\right> = \sigma\, T_{d1}^4 \, \left<\kappa_{abs}(a,T_d(R_1))\right>  \ .
\end{equation}

The upper limit on the dust mass determined by the IRAC upper limits is then given by
\begin{equation}
\label{mdup}
M_d^{{\rm up}} = {(\xi-1) \pi D^2\, F_{\nu}^{{\rm up}}(\lambda) \over  \int_{\xi}^1\ \kappa_{abs}(a,\lambda)\, \pi B_{\nu}[\lambda, T_d(x)]\, {\rm d}x } \ .
\end{equation}

 \subsection{The Effect of Optical Depth}
\begin{figure*}[t]
\begin{center}
\includegraphics[width=3.2in]{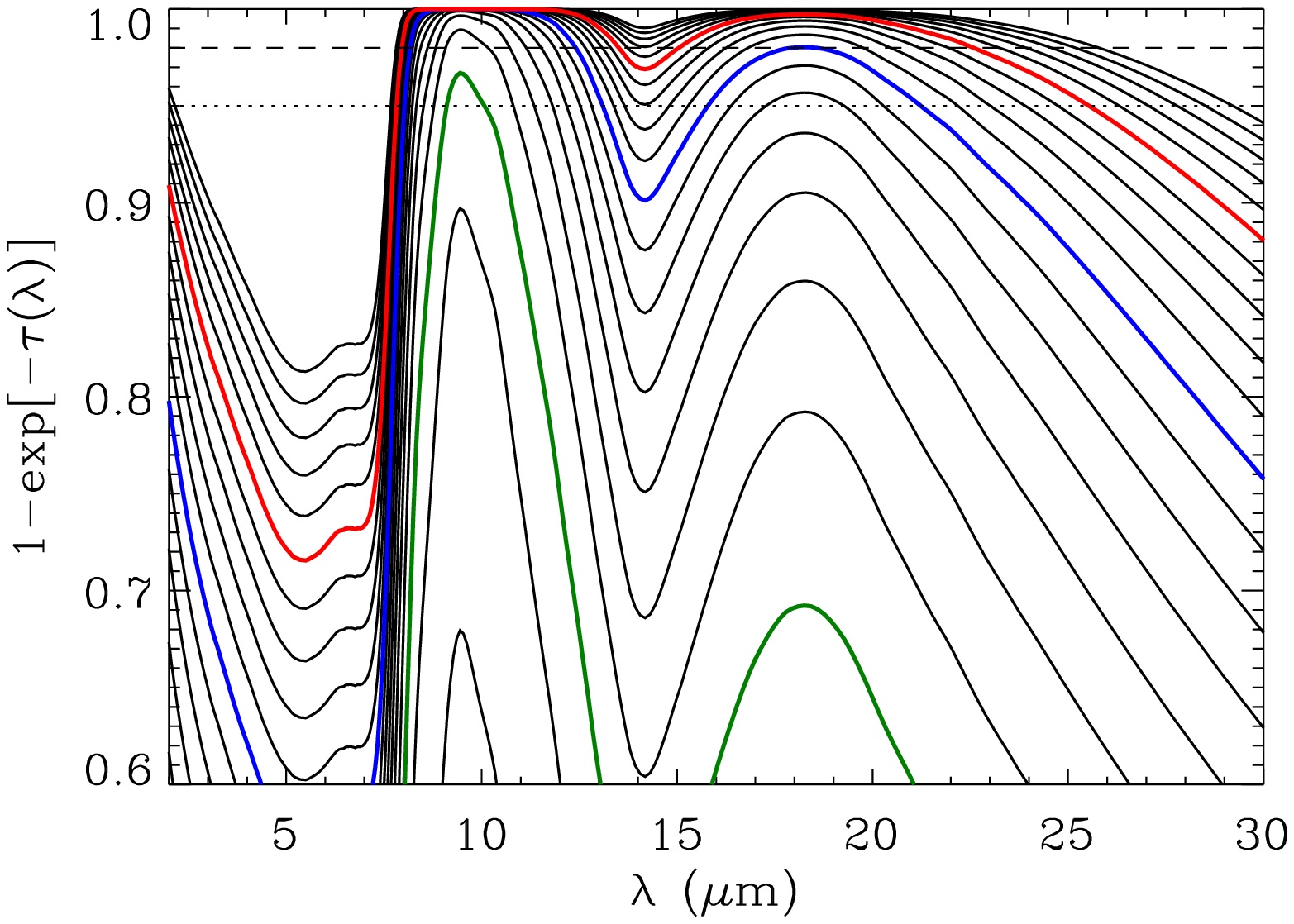}
\includegraphics[width=3.2in]{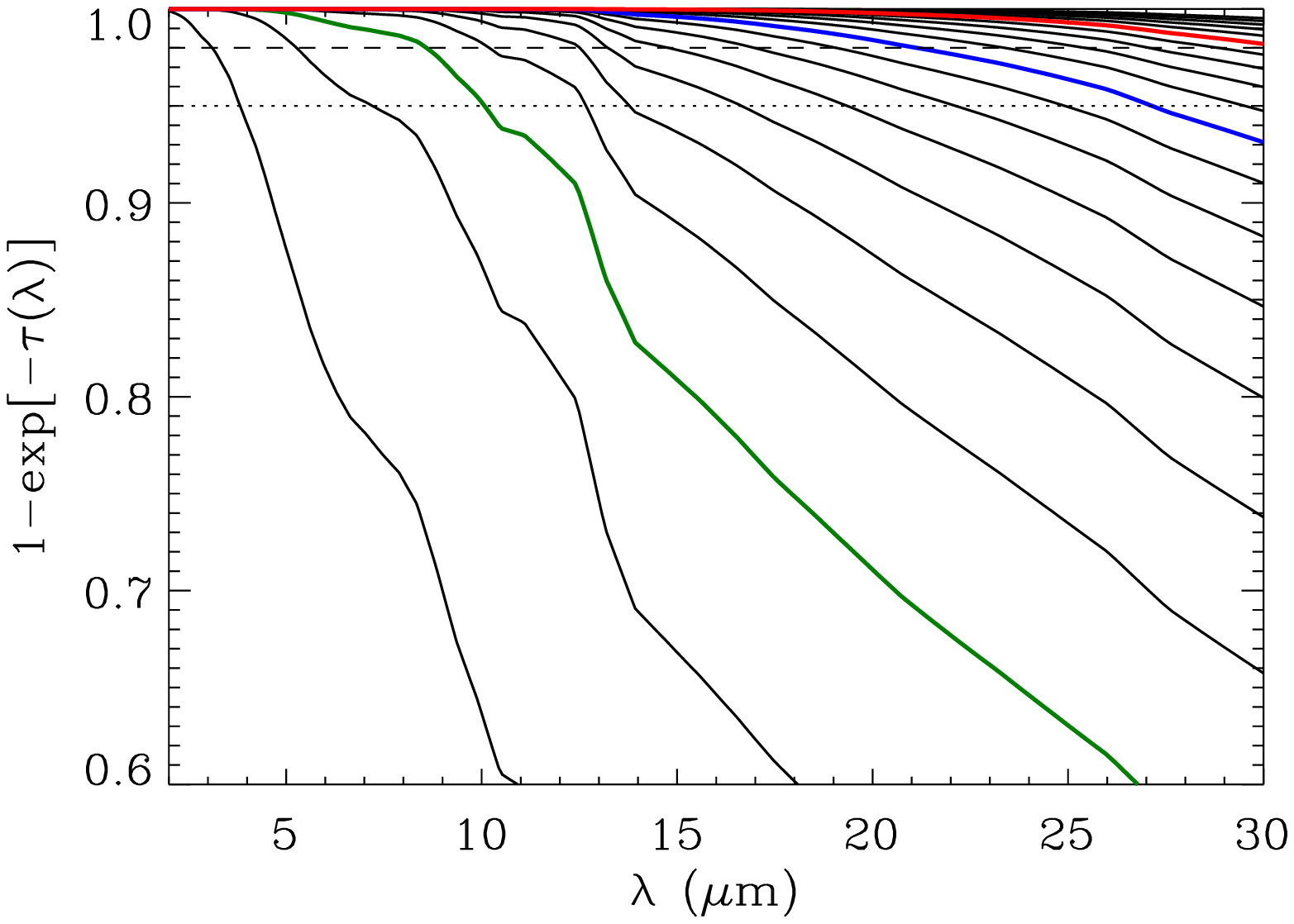}
\caption{\label{pesc} The ratio of the emerging spectrum from a slab of silicate (left panel) and ACAR (right panel) dust to that of a blackbody, given by Eq.~(\ref{slab}), as a function of wavelength for different values of the 10~\mic\ optical depth. The green, blue, and red curves correspond to $\tau(10~$\mic) values of 3, 10, and 15, respectively. The dotted and dashed horizontal lines correspond to 3 and 5\% deviations of the spectrum from that of a blackbody. The figure shows the 10~\mic\ optical depth beyond which it is impossible to determine the dust mass or composition.}
\end{center}
\end{figure*}

 The upper limit on the dust mass was derived under the assumption that the CSM is optically thin to its own radiation. 
 A fundamental limit on the ability to extract information on the CSM dust mass or composition is reached when the medium becomes optically thick at the wavelengths of observations.
 We present here a simple estimate of the optical depth at which the emerging IR spectrum becomes indistinguishable from that of a blackbody. 
The brightness of a radiating slab of dust with an optical thickness $\tau(\lambda)$ is given by
\begin{equation}
\label{slab}
I_{\nu}(\lambda) = B_{\nu}(\lambda, T)\, \left[1-\exp[-\tau(\lambda)]\right] \ .
\end{equation}
Figure~\ref{pesc} depicts the value of $ I_{\nu}(\lambda)/B_{\nu}(\lambda, T)$ ratio as a function of wavelength, for a slab composed of silicate or ACAR dust,  for different values of the 10~\mic\ optical depth. The green, blue, and red curves correspond to $\tau(10 $\mic) values of 3, 10, and 15, respectively. Given the accuracy of the flux measurements, the figures show the limiting optical depth above which the observed spectrum becomes indistinguishable from that of a blackbody, preventing any determination of the dust mass or composition.  

For example, the fractional uncertainties in the 3.6 and 4.6~\mic\ \spitz\ observations of SN~2010jl \citep{williams15} are about 20 to 30\%. The 3.6 and 4.5~\mic\ emission from the CSM can therefore be fit with a blackbody function if the 10~\mic\ silicate optical depth exceeds a value of $\sim 13$, hiding the 9.7 and 18~\mic\ silicate features. In that case, the non~detection of the 9.7~\mic\ silicate feature cannot be taken as evidence for the absence of silicate dust in the CSM.

Using Eq.~(\ref{mdmin}) we can write the mass of dust needed to make the CSM thick at near- and mid-IR wavelengths as
\begin{eqnarray}
\label{mdthick}
M_d^{thick} & = & {4 \pi R_1^2\over \xi}\ \left[{\tau_{abs}^{thick}(10\mu m) \over \kappa_{abs}(10\mu m)}\right] \\ \nonumber
 &= & {3.3\times10^{-3}\,  R_{16}^2\over \xi}\ \  {\rm M}_{\odot} \ \  {\rm  silicate} \\ \nonumber
  &= & {1.6\times10^{-2}\,  R_{16}^2\over \xi}\ \  {\rm M}_{\odot}\ \  {\rm  ACAR} \ ,
\end{eqnarray}
where the numerical values were calculated for \\$\tau_{abs}^{thick}(10~\mu$m)=15, and where  $\kappa_{abs}(10~\mu$m)=2900 and 580~cm$^2$~g$^{-1}$ for silicate and ACAR grains, respectively. 

\begin{figure*}[t]
\begin{flushright}
\includegraphics[width=3.2in]{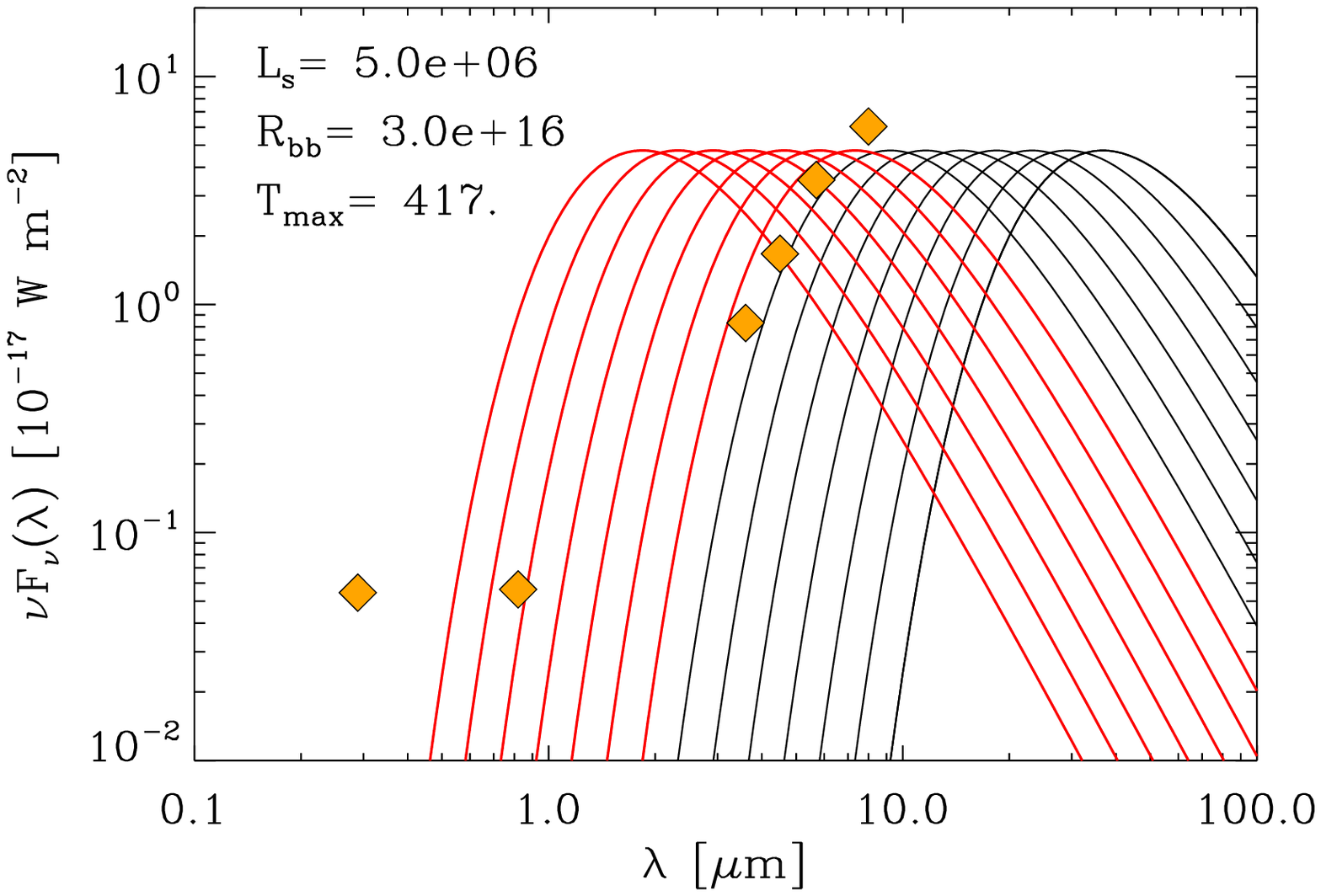}    
\includegraphics[width=3.2in]{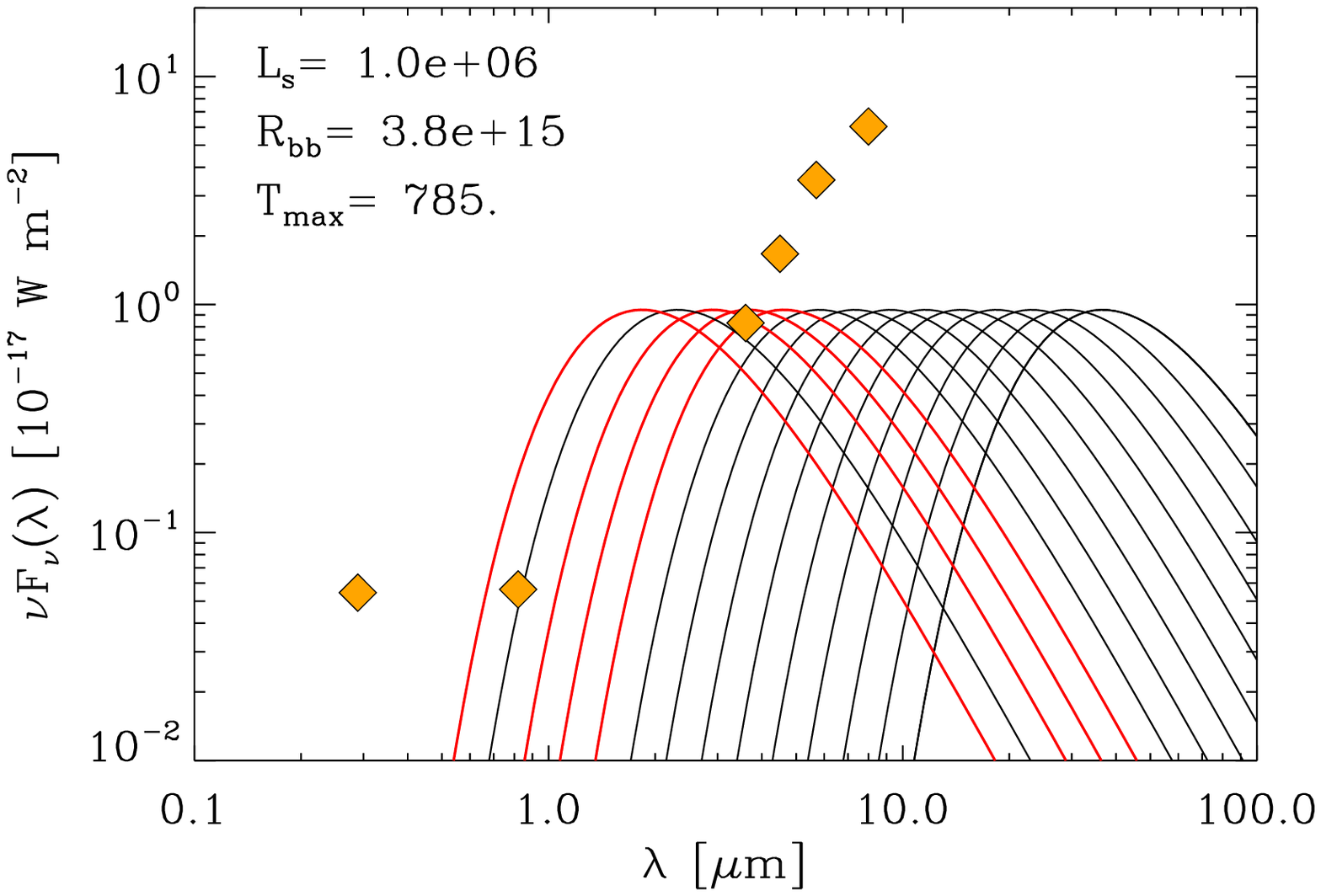} \\
\includegraphics[width=3.2in]{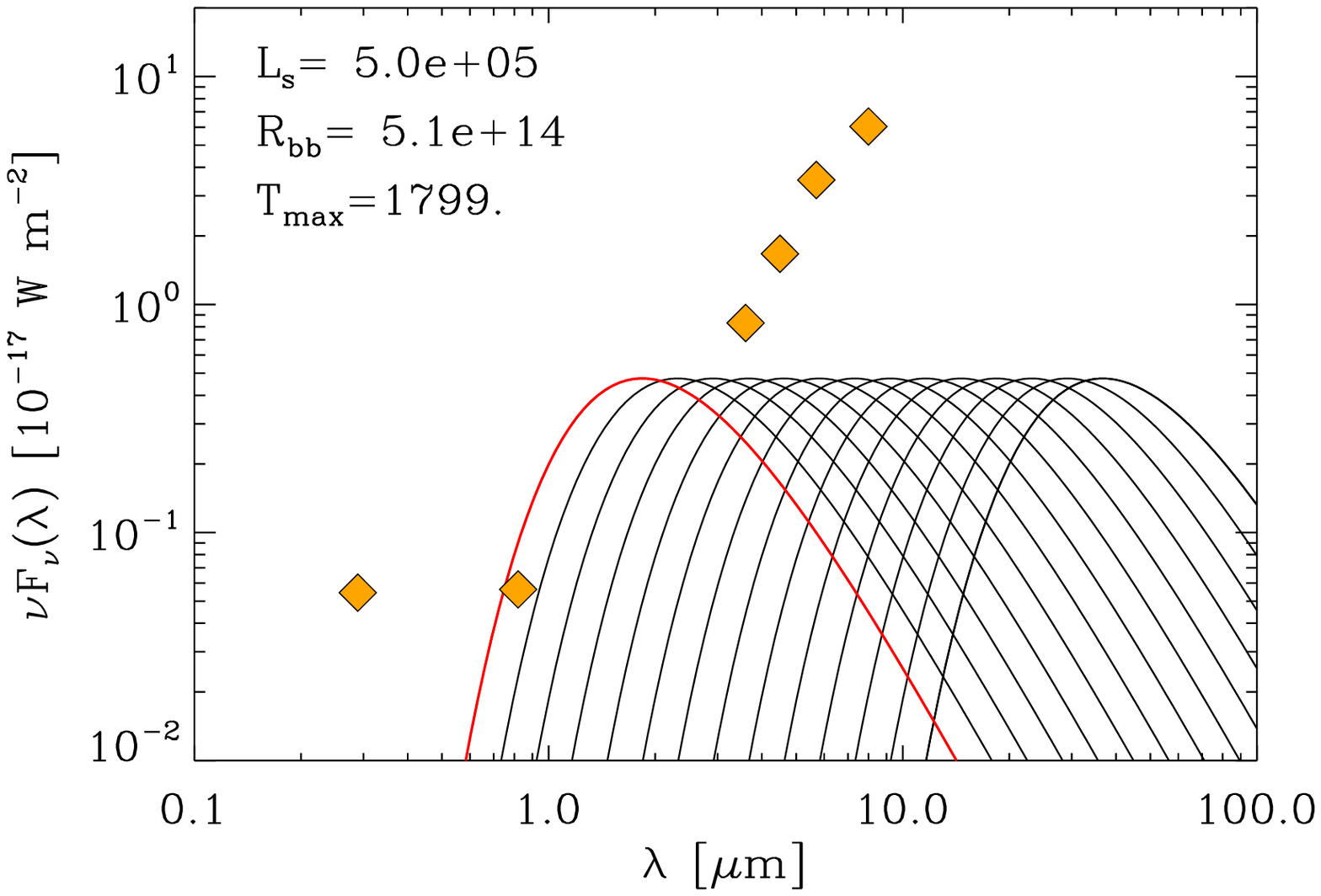} 
\includegraphics[width=3.2in]{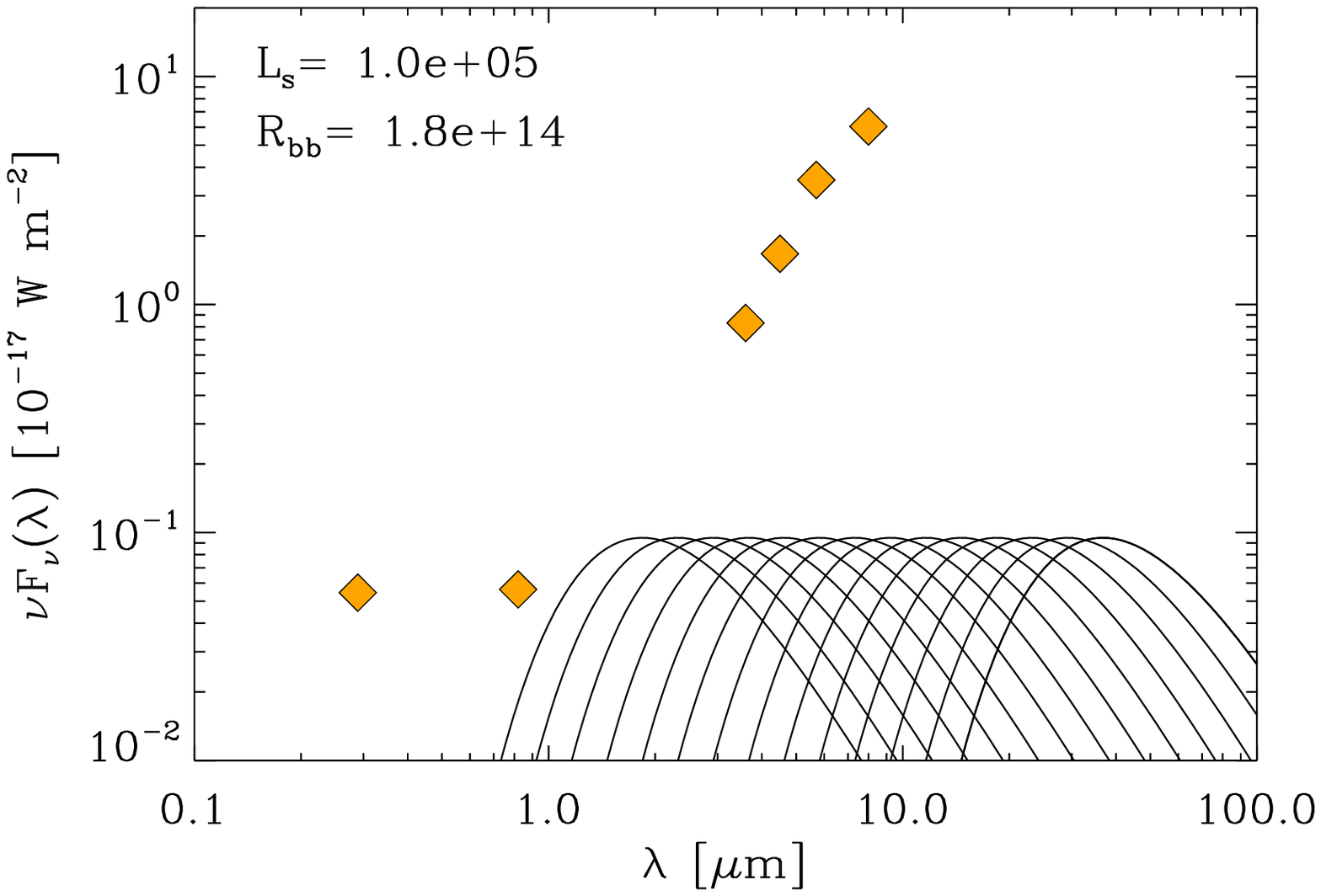} 
\caption{\label{bb} Blackbody spectra with temperatures ranging from 100 to 2000~K and normalized to the luminosity of the progenitor star are compared to the observed UVNIR upper limits. Red curves correspond to spectra that violate the observed upper limits. Each panel lists the maximum allowed blackbody temperature and the minimum blackbody radius of the CSM. 
}
\end{flushright}
\end{figure*}

As the CSM spectrum approaches that of a blackbody, its radius can be derived from the relation $4\pi\, R_{bb}^2\, \sigma\, T_{bb}^4 = L_s$, where $\sigma$ is the Stephan-Boltzmann constant, $T_{bb}$ is the CSM temperature, and $L_s$ is the luminosity of the progenitor star. Figure~\ref{bb} depicts the blackbody spectra of CSM surrounding progenitors with luminosities of $5\times 10^6$, $1\times 10^6$, $5\times 10^5$, and $1\times 10^5$~\lsun, for different CSM temperatures, ranging from 100 to 2000~K. For the three most luminous progenitors there is a maximum temperature above which the CSM spectrum will exceed the limits imposed by the UVNIR observations. This maximum temperature defines a minimum blackbody radius for the CSM. For the lowest luminosity progenitor, the CSM spectra never exceed the UVNIR limits even for the highest possible temperature of 2000~K. The minimum blackbody radius is below $10^{15}$~cm. So as the CSM becomes thick to its own emission, its radius cannot be smaller than the blackbody radius at the maximum allowed shell temperature.       

\section{Limits on the dust mass and distance for different stellar progenitors}

\begin{figure*}[t]
\begin{flushright}
\includegraphics[width=3.2in]{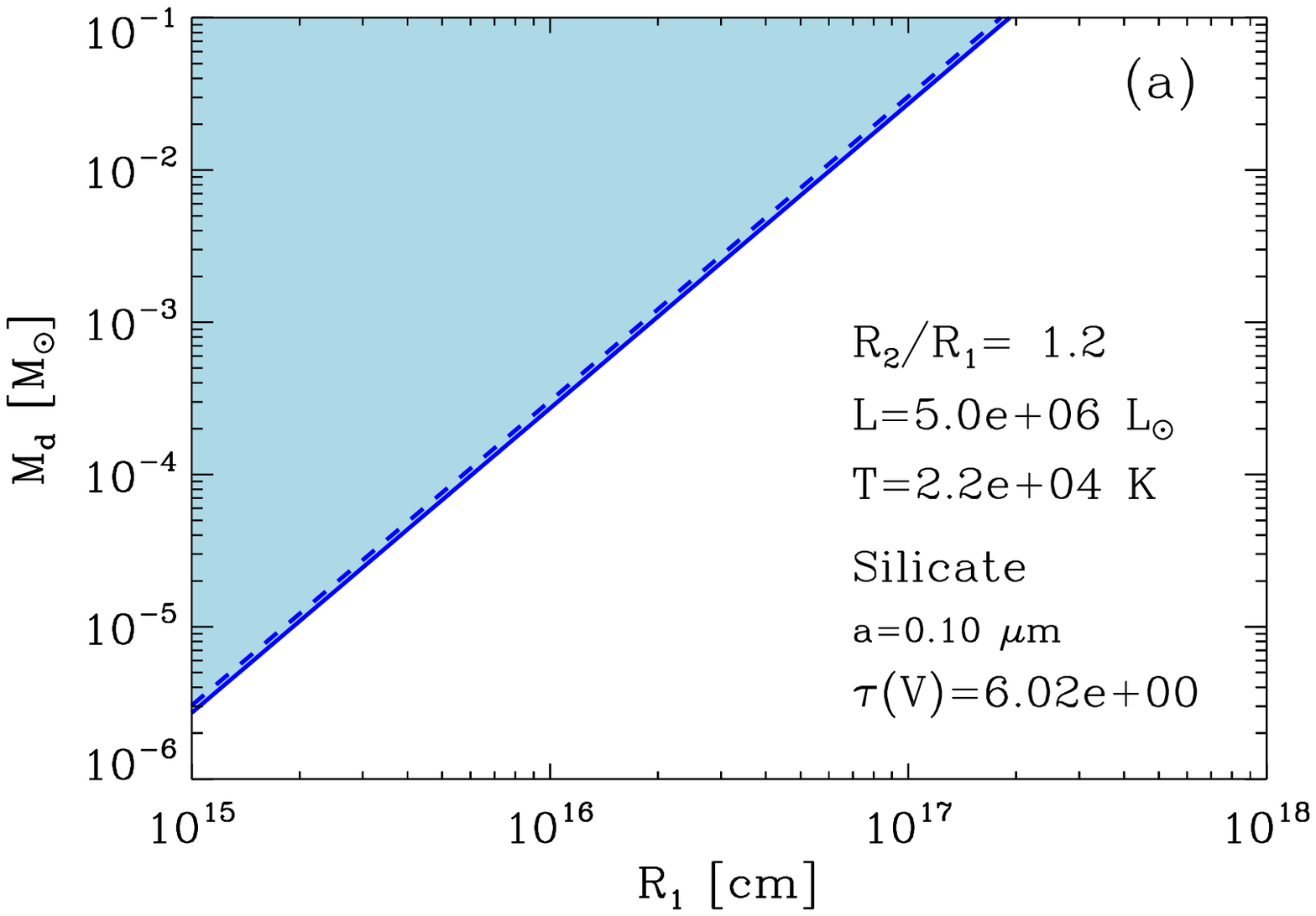}    
\includegraphics[width=3.2in]{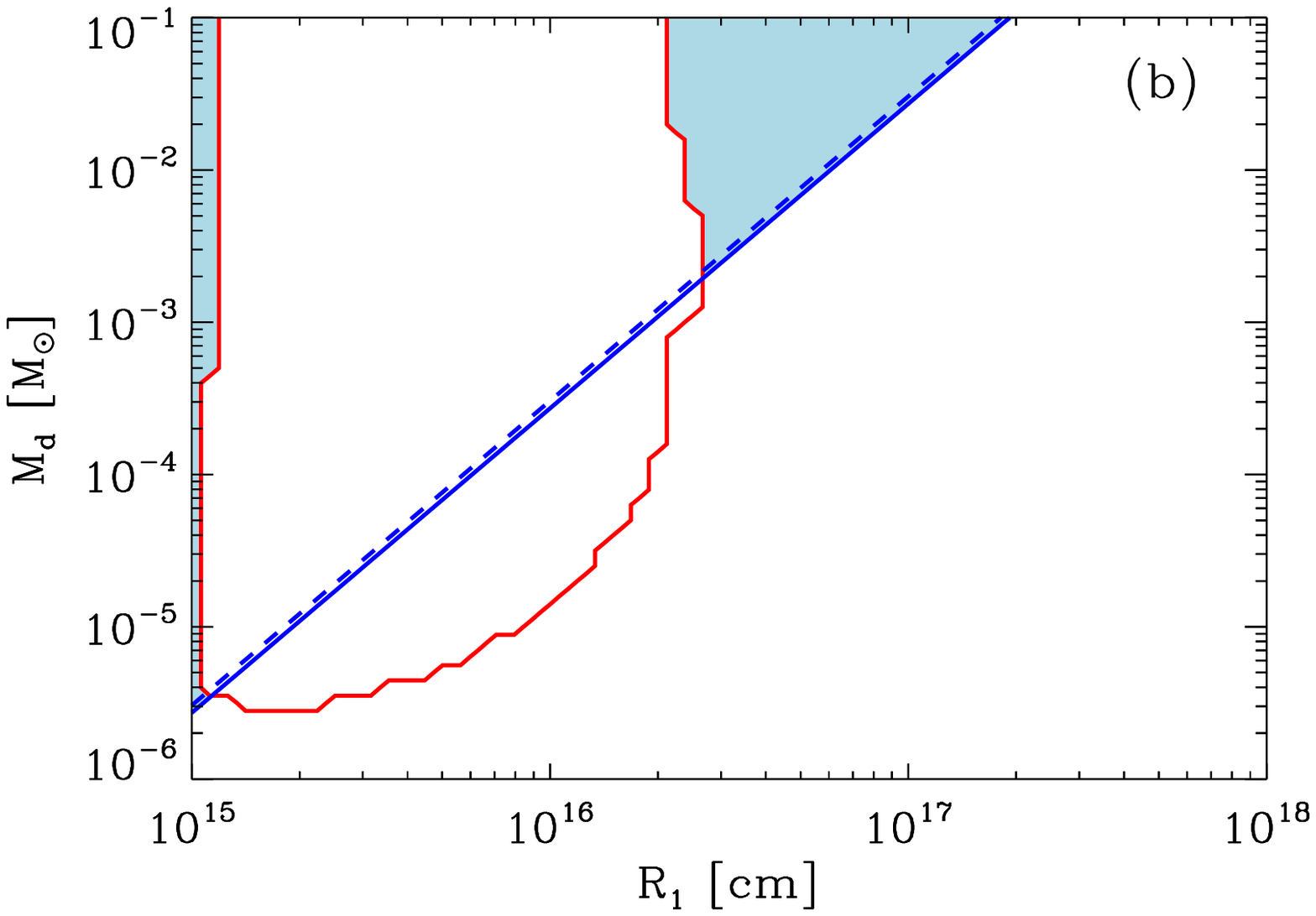} \\
\includegraphics[width=3.2in]{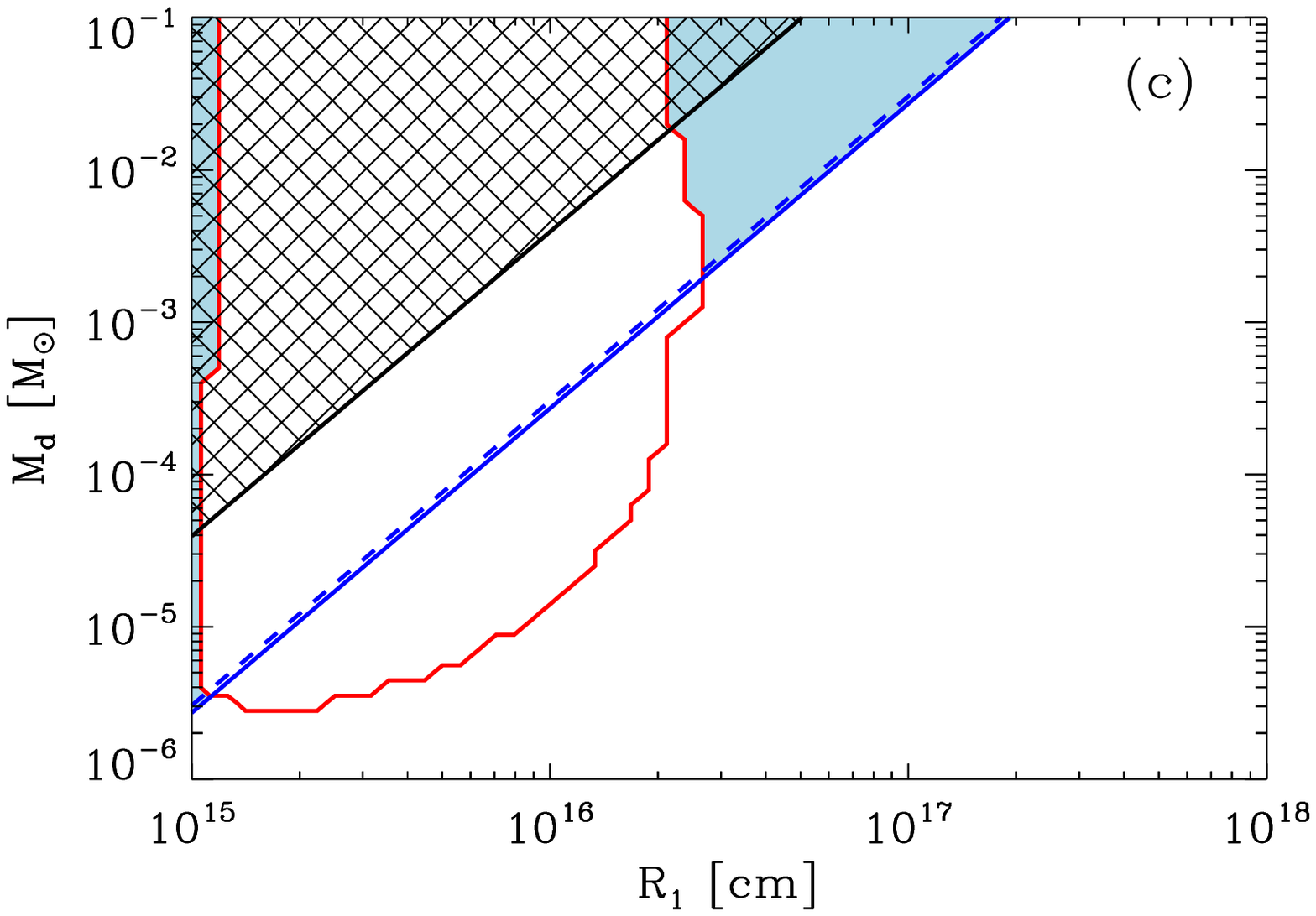} 
\includegraphics[width=3.2in]{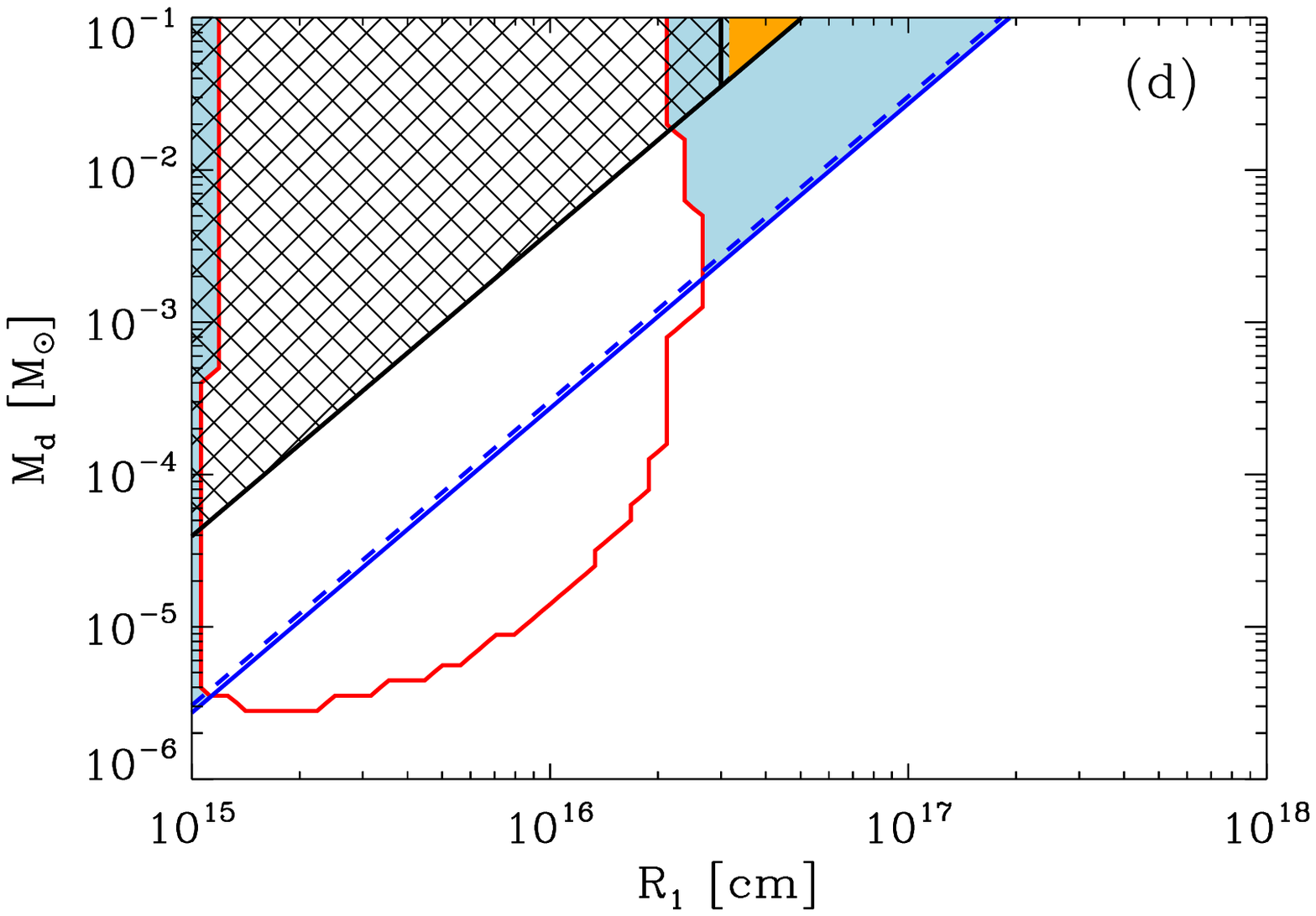} 
\caption{\label{build} An illustrative figure depicting the effects of the various observational constraints on the relation between the CSM dust mass and its inner radius, for an $R_2/R_1$ ratio of 1.2. The shell consists of 0.1~\mic\ silicate grains,  and surrounds a hot LBV progenitor that is characterized by a luminosity of $5\times 10^6$~\lsun and a temperature of 22,000~K.  The figure also list the value of $\tau(V)$, the minimum amount of visual extinction required to hide the progenitor star. (a) The blue shaded area reflects the allowed region that complies with the extinction requirement. The dashed blue line represents the minimum CSM dust mass required to hide the progenitor star, given by Eq.~(\ref{mdmin}), and the solid blue line represents the lower CSM dust mass required, when a LOS extinction of $\tau_{LOS}(V)=0.64$ is taken into account [Eq.~(\ref{mdminLOS})]. When the required extinction $\tau(V)$ falls below the LOS value, as is the case for low-luminosity ($L \leq 1\times 10^5$~\lsun) progenitors, all masses and radii depicted in the figure are allowed. (b) The red contour encloses the region in which the CSM flux exceeds the IRAC upper limits. $M_d-R_1$ combinations within this region are therefore not allowed. (c) The cross-hatched region corresponds to the cases in which the 10~\mic\ optical depth of the CSM exceeds 15, and its spectrum approaches that of a blackbody. Consequently, the possibility to retrieve any information on either the mass or the composition of the CSM  depends on the accuracy of the flux measurements.  (d) The orange shaded area depicts the regions of CSM shells that are characterized by  blackbody spectra that do not violate the UVNIR upper limits. The CSM radius must therefore exceed that of a blackbody with a luminosity equal to that of the embedded progenitor star and radiating at the highest allowable temperature (see Fig~\ref{bb}). Allowed regions are therefore those shaded in blue or orange. More details are in the text.
}
\end{flushright}
\end{figure*}

Figure~\ref{build}  illustrates the cumulative effects of the different observational and physical constraints on the permissible  $M_d-R_1$ relation for a CSM shell characterized by an $R_2/R_1$ ratio of 1.2, comprising of 0.1~\mic\ silicate grains. The progenitor is a hot LBV star with a luminosity and temperature of  $5\times 10^6$~\lsun\  and 22,000~K, respectively. 

The blue shaded area in Figure~\ref{build}(a) depicts the permissible $M_d-R_1$ combination imposed by the requirement that the CSM optical depth be high enough to hide the progenitor star. The dashed blue line represents the dependence of the minimum CSM dust mass, $M_d^{\rm min}$, on $R_1$ in the absence of any LOS extinction, given by eq.~(\ref{mdmin}.  The mass of CSM dust required to hide the progenitor is reduced in the presence of LOS extinction, and the thick blue line represents the  $M_d^{\rm min}-R_1$ relation for a LOS extinction of $\tau_{LOS}(V)=0.64$, [see Eq.~(\ref{mdminLOS})].

The red contour in Figure~\ref{build}(b) encloses the region in which the IR emission from the CSM exceeds the IRAC upper limits. $M_d-R_1$ combinations within this region are therefore not allowed.

The cross-hatched region in Figure~\ref{build}(c) corresponds to the cases in which the 10~\mic\ optical depth of the CSM exceeds 15, and its spectrum approaches that of a blackbody. Consequently, the possibility to retrieve any information on either the mass or the composition of the CSM  depends on the accuracy of the flux measurements.
As the CSM spectrum approaches that of a blackbody, its radius must exceed the blackbody radius corresponding to the progenitor's luminosity and the maximum temperature allowed by the UVNIR limits shown in Figure~\ref{bb}. In other words, energy conservation dictates that the IR emission from the CSM must equal that of the progenitor star. Dust at close proximity to the progenitor star will therefore have to radiate at a high temperature to satisfy the energy constraint, and consequently violate the UVNIR upper limits.  

The orange shaded area in Figure~\ref{build}(d) is bounded at the lowest radius by the blackbody radius presented in Figure~\ref{bb}, and depicts the regions of CSM shells that are characterized by blackbody spectra that do not violate the UVNIR upper limits. 

In summary, the only allowable dust masses and regions are those presented by the clear blue and orange regions, with the caveat that no information on the dust mass or composition can be retrieved from the latter.   

\begin{figure*}[t]
\begin{center}
\includegraphics[width=3.0in]{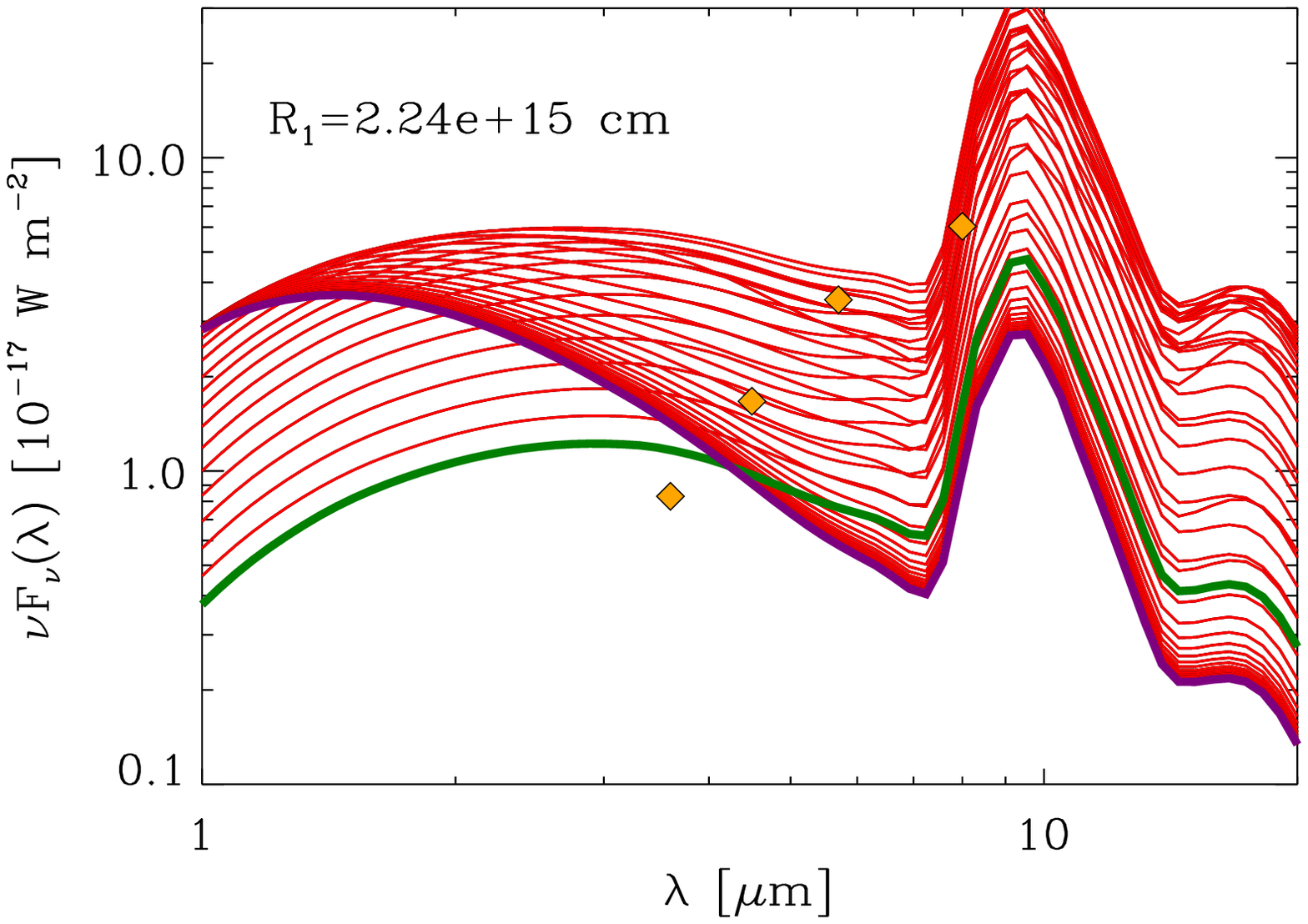}   
\includegraphics[width=3.0in]{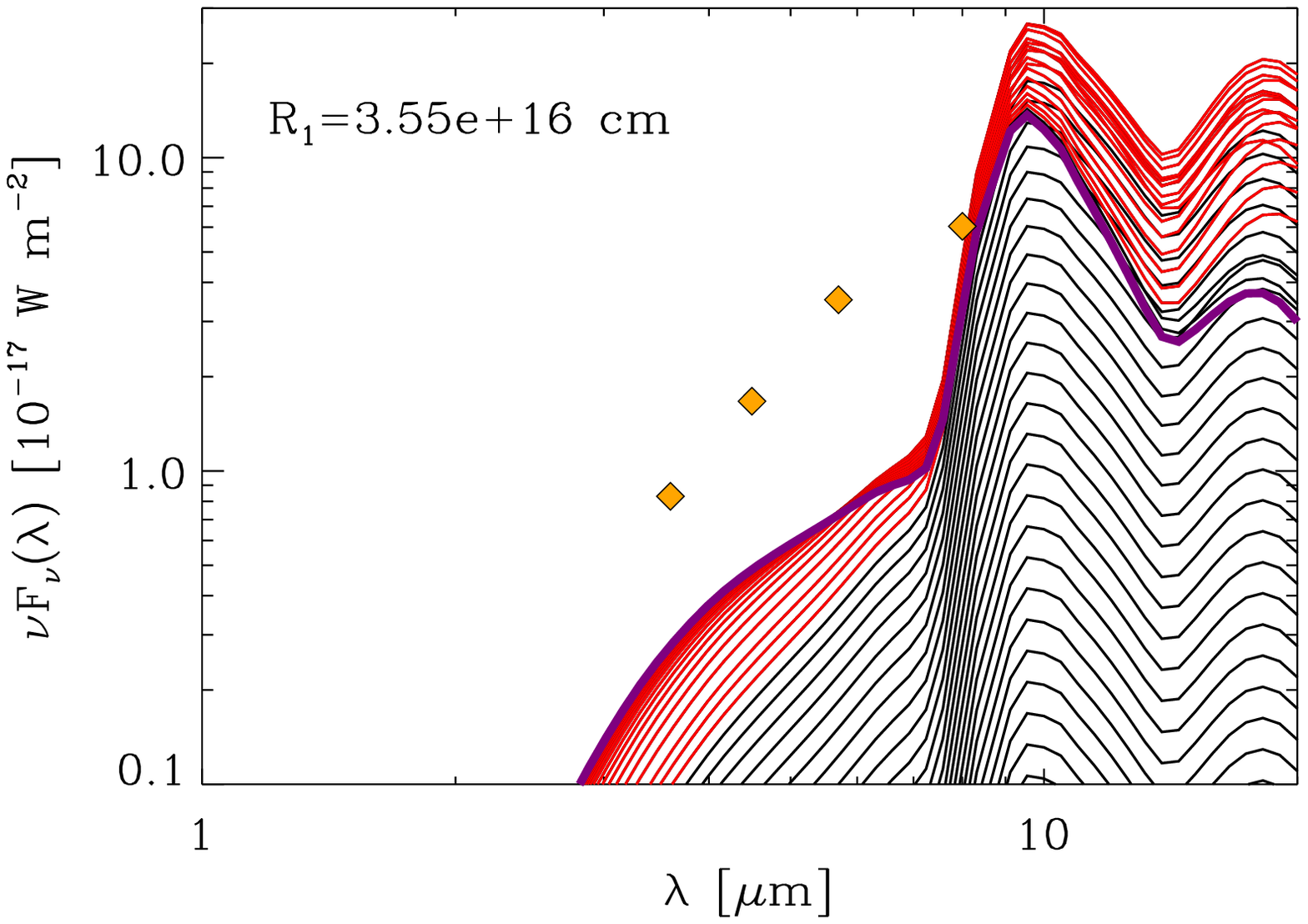}\\
\includegraphics[width=3.0in]{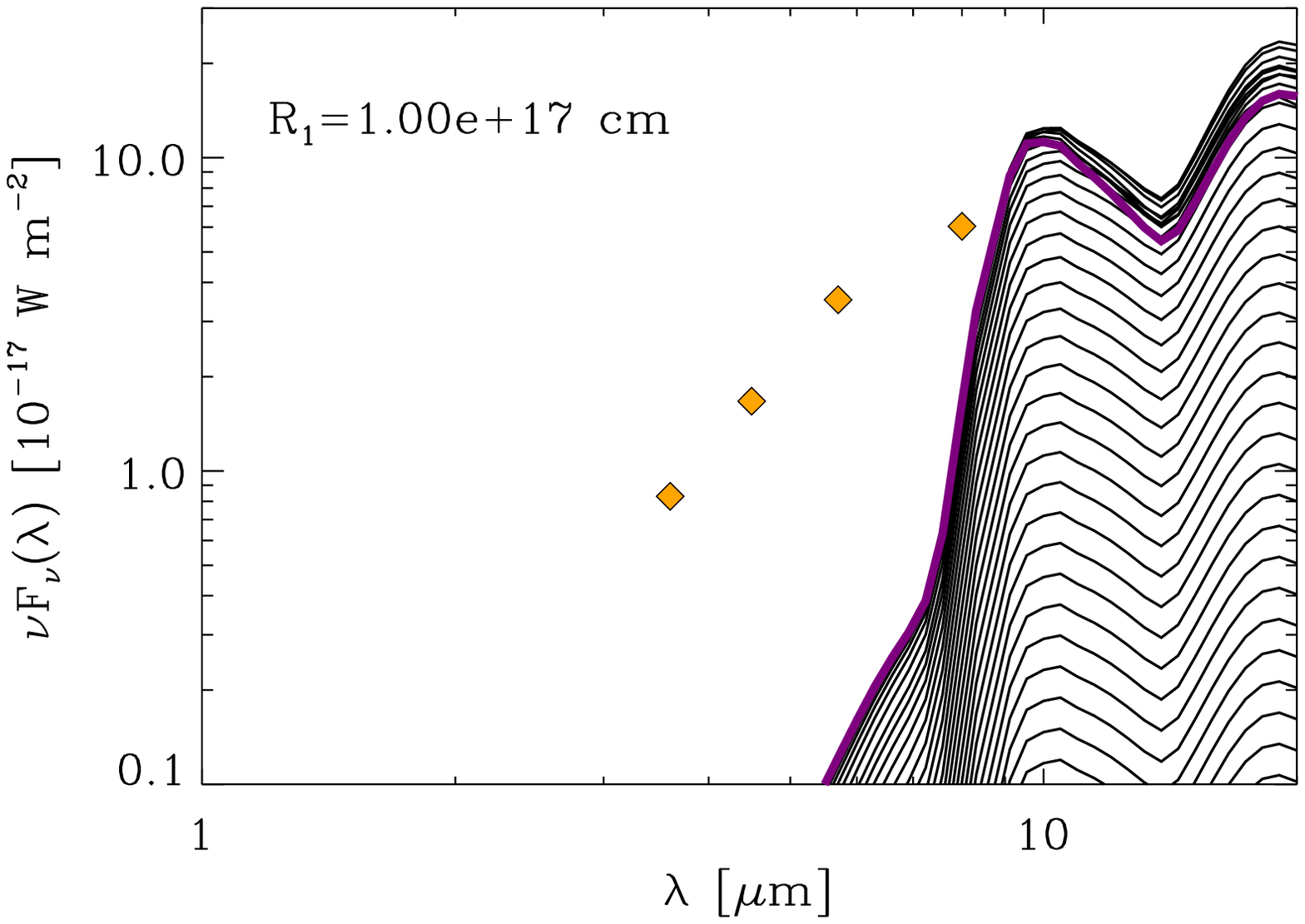} 
\includegraphics[width=3.0in]{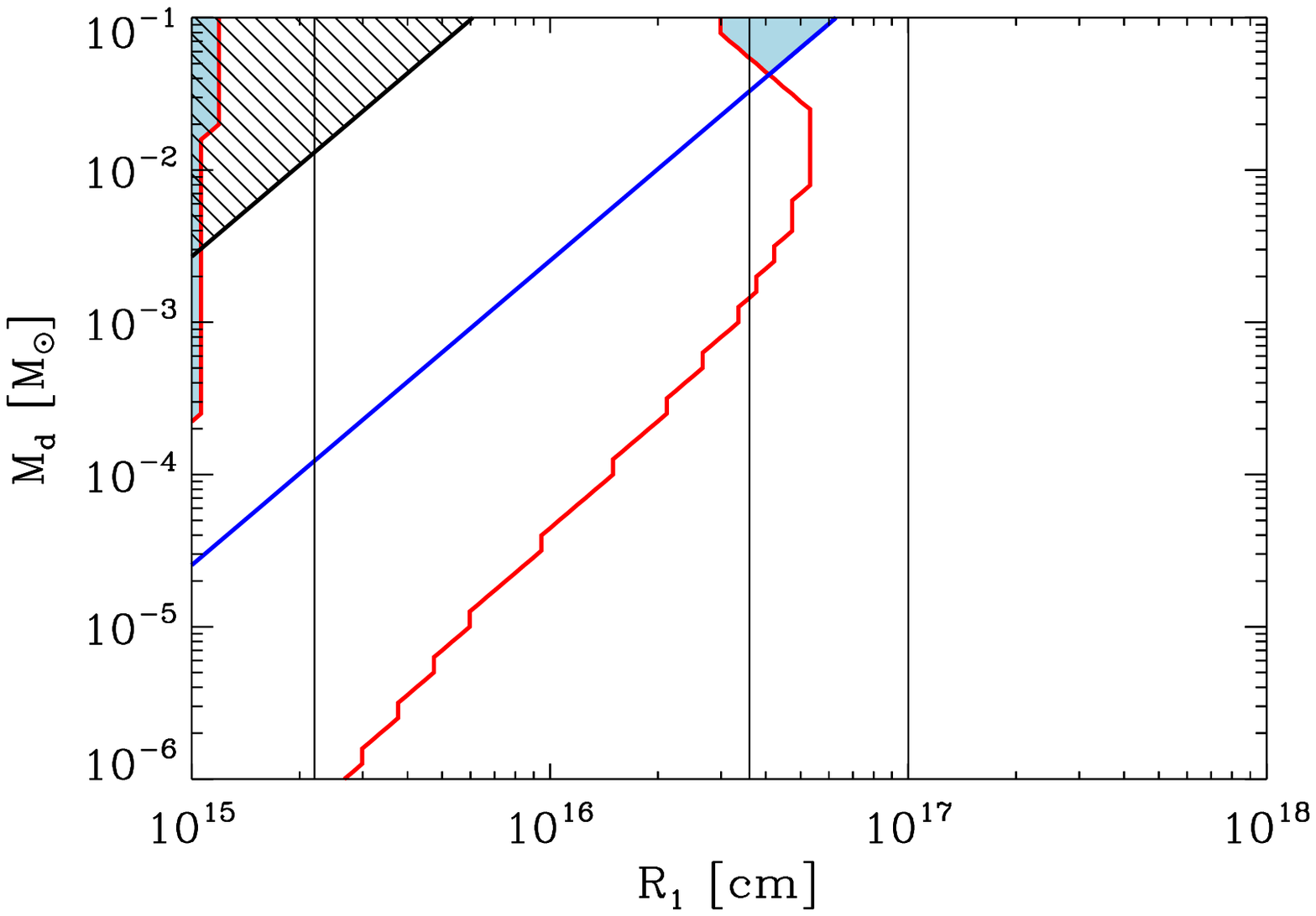}
\caption{\label{spec} {\bf Top row and bottom-left panels}: Spectra of 0.1~\mic\ silicate grains heated by a hot LBV with a luminosity and temperature of $5\times 10^6$~\lsun\ and 22,000~K, respectively, for CSM with inner radii of $R_1 = 2.0\times 10^{15}, 2.0\times 10^{16}$, and $1.0\times 10^{17}$~cm. The spectra correspond to different CSM dust masses. Red spectra exceed the IRAC upper limits. The green and purple curves depict the spectra for masses of  $1\times 10^{-6}$ and 0.1~\msun, respectively.  {\bf Bottom-right panel}: The $M_d-R_1$ diagram, taken from Figure~\ref{sil1} corresponding to the model parameters of the spectra. The vertical lines correspond to the positions of the inner radii.
}
\end{center}
\end{figure*}

Figure~\ref{spec} depicts the spectra of CSM surrounding a hot LBV progenitor with a luminosity and temperature of $5\times 10^6$~\lsun and 22,000~K, respectively. The CSM consists of 0.1~\mic\ silicate grains. The spectra represent those of CSM with inner radii of $R_1 = 2.2\times 10^{15}, 3.6\times 10^{16}$, and $1.0\times 10^{17}$~cm, and different CSM masses. Red spectra correspond to masses that exceed the IRAC upper limits. The bottom-left panel in the figure shows the $M_d-R_1$ diagram for this case (see Figure~\ref{sil1}). 

When $R_1 = 2.2\times 10^{15}$~cm, the dust is too close to the central star, and all spectra violate the NIR limits. Further out, when $R_1=3.6\times 10^{16}$~cm, some spectra (shown in red) for which the dust masses lie within the red contour lines are forbidden, while those corresponding to CSM dust masses below $\sim 10^{-3}$~\msun\ or above $\sim 0.04$~\msun\ are allowed. At masses below $\sim 10^{-3}$~\msun\ the emission is not high enough to violate the NIR upper limits, while at large ($\gtrsim 0.04$~\msun) masses, the shell becomes partially thick to the stellar radiation, so that the emission from the hot dust does not exceed the NIR upper limits. 

At $R_1=1.0\times 10^{17}$~cm, the CSM shell is so cold that none of the spectra exceeds the NIR upper limits. However, because of the large inner radius of the CSM, the dust mass surface density, and consequently the optical depth, is too low to provide the necessary extinction to hide the progenitor star. A dust mass of $\sim 1$~\msun\ is then required to provide the necessary extinction. Such large dust masses would require an unlikely scenario of an extremely massive progenitor which has lost over 100~\msun\ of material over its main-sequence lifetime.

The figure illustrates several important points. (a) The NIR upper limits can only constrain the amount of hot CSM dust. The mass of cold dust  is constrained by the total mass of the CSM gas that follows the $r^{-2}$ density profile. The CSM dust mass will probably not exceed 0.1~\msun, if the total mass lost by the progenitor star is $\lesssim 10$~\msun. (b) The NIR upper limits cannot constrain any amount of cold dust that is not part of the assumed quiescent mass loss.  (c) All CSM spectra were calculated assuming that the reradiated IR emission escapes the shell. This is obviously not the case when $R_1 = 2.2\times 10^{15}$~cm and $M_d \gtrsim 10^{-3}$~\msun, when the 10~\mic\ optical depth of the shell exceeds 15. The emerging IR spectrum should then be close to that of a blackbody. However, this approximation does not affect the conclusion that all the spectra at this radius  violate the NIR upper limits.   

Figure set~\ref{sil1}  combines all constraints on the CSM dust mass as a function of its inner and outer radii, $R_1$ and $R_2$, for different types of progenitor stars, dust composition, and grain radii. 
The figures are grouped for different  progenitors and dust compositions. The columns in each figure correspond to different values of $R_2/R_1$, and the rows to different grain radii. Each panel in the figures also lists the visual optical depth required to hide the progenitor star. An optical depth of $\tau(V)=0.64$ is provided by foreground extinction.

The results for hot LBV progenitors, cool LBVs, YSG, and RSG progenitors are presented in figure set~\ref{sil1}. 

The IRAC upper limits constrain the mass and location of the dust primarily for hot and cool LBVs with luminosities of $5\times10^6$~\lsun for silicates and $> 5\times10^5$~\lsun\ for ACAR grains. Typical silicate masses and CSM proximities to the progenitor stas are $\gtrsim 10^{-3}$~\msun\ and $\gtrsim 10^{16}$~cm, respectively. Because of their higher absorptivity,  a CSM consisting of ACAR dust must be at larger distances from a hot LBV than a CSM consisting of silicate dust. 
IRAC upper limits provide some constraints on the mass and location of ACAR dust around the hotter YSG and RSG progenitors. For all other cases, the extinction requirements provide the only constraint on the location and dust mass of the CSM.

\section{Discussion}

Post-explosion X-ray, UV-optical, and NIR observations of SN~2010jl can shed further light on the presence of any pre-existing dust in its CSM. 
Post-explosion X-ray observations of SN~2010jl required an obscuring CSM H~column density of $N_{\rm H} \approx (0.1-1)\times 10^{24}$~cm$^{-2}$ to fit to the X-ray data at the first epoch of X-ray observations, when the shock radius was about $10^{15}$~cm \citep{ofek14,chandra15}. In general, the dust optical depth at wavelength $\lambda$ can be written as
\begin{equation}
\tau(\lambda) = Z_{d{\rm H}}\, m_{\rm H}\, \kappa_{ext}(\lambda)\, N_{\rm H}\ ,
\end{equation}
where $m_{\rm H}$ is the hydrogen mass, and $Z_{d{\rm H}}$ is the dust-to-H mass ratio.  
For a value of $\kappa_{ext}({\rm V}) \approx 10^4$~cm$^2$\, g$^{-1}$ and $N_{\rm H} = 10^{23}$~cm$^{-2}$, we get  $\tau({\rm V}) \approx  1.7\times10^3\, Z_{d{\rm H}}$. So the formation of a modest amount of dust in the progenitor wind, characterized by a dust-to-H mass ratio of $\sim 6\times10^{-4}$ [the Galactic ratio is 0.007 \citep{zubko04}], will lead to a visual extinction of $\sim 1$. 

Post-explosion data suggest that the progenitor likely underwent one or more eruptions, during which the majority of the mass loss occurred \citep{andrews11, maeda13,fox13b,ofek14,chandra15}.  These studies place a shell of material at $10^{16}\lesssim R_1 \lesssim 10^{17}$~cm with a dust mass of $M_d \approx 10^{-2}$~\msun, accommodating many scenarios considered in this study. 

Additional constraints on the structure of the CSM may be inferred from X-ray observations of SN~2010jl. The X-ray flux decreases initially as $t^{-0.38}$ before, and as $t^{-3.14}$ after, $t\approx 350$~d \citep{ofek14}. The initial decline of the X-ray flux, and the decrease in the obscuring H~column density \citep{chandra15}, are consistent with a shock wave that is propagating through a CSM with an $r^{-2}$ density profile. The break in the X-ray flux, which is also observed at UV/optical wavelengths, is likely the result of the shock reaching the outer boundary of one of the shells, which yields an outer radius of $\lesssim 3\times 10^{16}$~cm \citep{ofek14, chandra15}. Such a constraint will rule out a CSM consisting of ACAR grains surrounding a hot LBV.

We assumed throughout the paper that the CSM is spherical in nature. Andrews et al (2011) concluded that the post-explosion IR emission from the CSM must arise from an inclined torus, that does not intercept the line of sight, based on the lack of extinction to the SN in the host galaxy.  Such configuration can only be valid if any preexisting dust needed to hide the progenitor star was either vaporized by the SN flash or destroyed by the SN blast wave. 
Additional evidence for asymmetric CSM comes from spectropolarimetry \citep{patat11} and from observations of slow and fast velocity component which may be due to  a bipolar CSM configuration  \citep{smith11}.  The low velocity component arises from the equatorial region, and the faster from a polar outflow.

Both an inclined toriodal or a bipolar CSM intersecting the LOS can provide the necessary extinction.  The implication of the UVNIR upper limits on the dust mass, dust properties,  and distance to the progenitor will then depend on the specific CSM geometry. 

Finally, further information  can be obtained from the evolution of the IR light curve of the SN \citep{gall14,fransson14}. The early IR emission may consist of the fraction of the SN's X-ray and UV emission that was reprocessed by CSM dust \citep[][; Dwek  et al. 2017, in preparation; Sarangi et al. 2017, in preparation]{andrews11,fransson14}. Analysis of this possible IR echo will provide further constraints on the composition, sizes, and  spatial distribution of any pre-existing CSM dust .

\section{Summary}
\begin{figure*}[t]
\begin{center}
\includegraphics{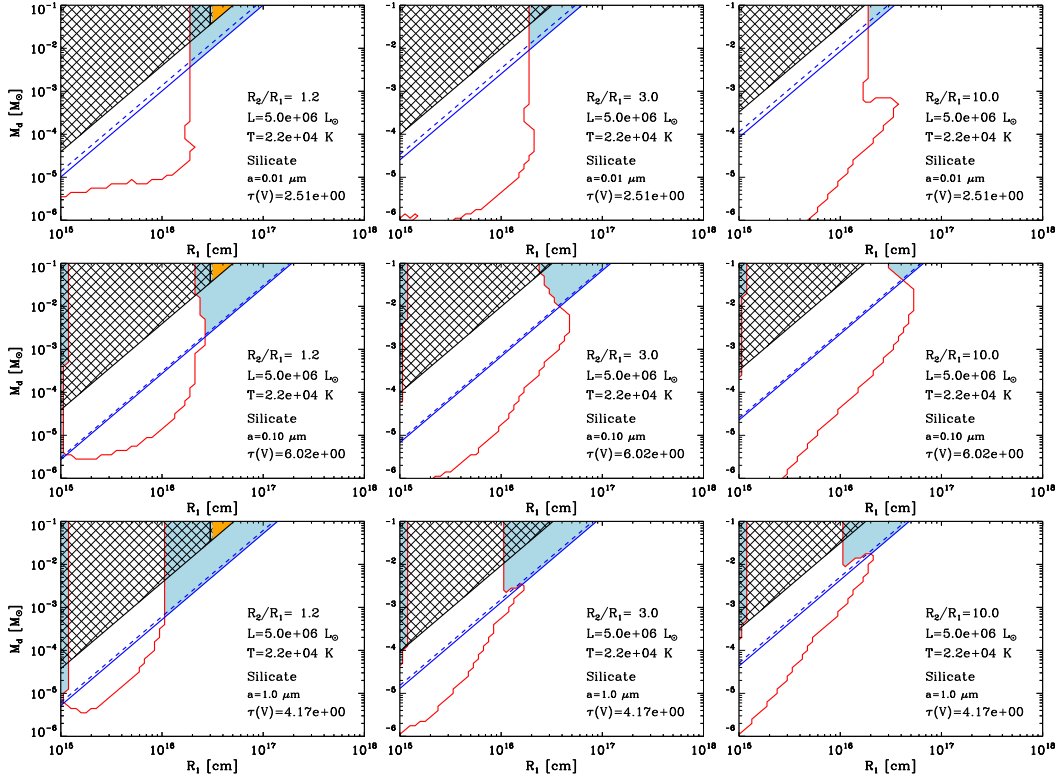}
\caption{\label{sil1} Limits on the mass of the CSM dust composed of silicate grains surrounding a hot LBV with a luminosity and temperature of $5\times 10^6$~\lsun and 22,000~K, respectively, as a function of the inner-shell radius $R_1$, for different values of $\xi^{-1} \equiv R_2/R_1$ and grain radii. 
The dashed blue line indicates the {\it lower} limit on the circumstellar dust mass needed to create sufficient extinction to keep the flux from the progenitor star below the observed UVNIR limits in the absence of any LOS extinction. The solid blue line represents the same quantity with a LOS extinction of $\tau_{LOS}(V)=0.64$. The red line encloses the region where the CSM dust mass is such that the IR flux from the shell exceeds the IRAC upper limits. Only $M_d-R_1$ combinations outside the contours are allowed by the IRAC upper limits. Blue shading indicates regions where the CSM dust can provide sufficient extinction to hide the progenitor without exceeding observed IRAC upper limits or physical limits on the reradiated luminosity. The black stripes indicate the region where the CSM becomes optically thick at 10~\mic, and no information can be extracted on either the mass or the dust composition from IR ($\lambda \lesssim 10$ \mic\ observations. The orange shaded area depicts the regions of CSM shells that are characterized by blackbody spectra that do not violate the UVNIR upper limits. Also listed in the figures is the visual optical depth needed to hide the progenitor in the UVNIR. (The complete figure set (24 images) is available in the online journal.)
}
\end{center}
\end{figure*}

We have used the pre-explosion UVNIR upper limits on the emission from SN~2010jl to derive constraints on the properties of its progenitor star, and on the dust mass, distance, and extent of its surrounding CSM.
The results of the paper are presented in figure set~\ref{sil1}, and can be briefly summarized as follows.
\begin{enumerate}
  \item We have combined the UVNIR and IRAC constraints to map out the allowable CSM dust masses and inner radii as a function stellar characteristics, and of the  dust composition, grain sizes, and outer radius of the CSM. A detailed description of the construction of these constraints is presented in Figure~\ref{build} and related discussion. In general, the upper limits do not rule out any progenitor star, and only constrain the dust mass and proximity of the CSM to the progenitor as a function of the stellar and dust properties.
  
  \item {\bf Hot and cool LBV progenitors}: The most luminous LBVs are only allowed if they are surrounded by $\gtrsim 10^{-3}$~\msun\ of dust located at distances $\gtrsim 10^{16}$~cm from the progenitor star. 
  A wider range of dust masses and CSM proximities is allowed for lower-luminosity stars. For the faintest stars, a LOS visual extinction of $\tau=0.64$ would be sufficient to hide the progenitor.
  \item {\bf Yellow and red supergiant  progenitors}:  Any yellow or red supergiant progenitors are allowed, and their CSM dust mass and CSM proximity is only determined by the UVNIR extinction constraints. As for the LBV progenitors, a LOS visual extinction of $\tau=0.64$ may be sufficient to hide the faintest progenitors.
  \item For each dust composition, size, and $R_2/R_1$ ratio, there exists an $M_d\, R_1^2$ limit, given by Equation~(\ref{mdthick}), where $\tau(15~\mu m)\gtrsim 15$, and the CSM shell becomes optically thick to its own radiation. No information can then be obtained on either the mass or composition of the CSM. Furthermore, there is a lower limit on the radius of such CSM, since below that limiting radius the CSM temperature will have to rise above the limit imposed by the NIR upper limits.    
\end{enumerate} 


  
\acknowledgements
 E.D. and R.G.A. acknowledge NASA's 16-ATP2016-0004 grant for support on this project.  
  A.V.F.'s group is grateful for generous financial assistance from the
Christopher R. Redlich Fund, the TABASGO Foundation,  NSF grant
AST-1211916, and the Miller Institute for Basic
Research in Science (UC Berkeley) 
Support was also provided by NASA through HST grants
GO-14149 and GO-14668 from the Space Telescope Science Institute,
which is operated by AURA, Inc. under NASA contract NAS 5-26555.
The work of A.V.F. was completed in part at the Aspen Center for Physics,
which is supported by NSF grant PHY-1607611; he thanks the Center
for its hospitality during the neutron stars workshop in June and
July 2017.

\bibliographystyle{aasjournal.bst}
\bibliography{ms_no_figset.bbl}

\end{document}